\pdfoutput=1 
\documentclass[useAMS,usenatbib]{mn2e}
\usepackage{subfigure,amsmath}
\usepackage{color}

\usepackage{graphicx}

\definecolor{purple}{rgb}{1,0,1}

\newcommand{\lcdm}{$\Lambda$CDM}
\newcommand{\hmpc}{$h^{-1}$Mpc}
\newcommand{\gmpc}{$h^{-1}$Gpc}
\newcommand{\hmsol}{\mbox{ } {h}^{-1}~{M}_{\odot}}
\DeclareMathOperator\erf{erf}

\bibpunct[; ]{(}{)}{;}{a}{}{,}

\title{The dark matter of galaxy voids}
\author[P.~M. Sutter et al.]
{
\parbox{\textwidth}{
{P.~M. Sutter}$^{1,2,3,4}$ \thanks{Email: sutter@iap.fr},
Guilhem Lavaux$^{1,2,5,6,7}$,
Benjamin D. Wandelt$^{1,2,4,8}$,
David H. Weinberg$^{3,9}$, and
Michael S. Warren$^{10}$
}
\vspace{0.4cm}\\
\parbox[c]{\textwidth}{
$^{1}$ UPMC Univ Paris 06, UMR7095, Institut d'Astrophysique de Paris, F-75014, Paris, France \\
$^{2}$ CNRS, UMR7095, Institut d'Astrophysique de Paris, F-75014, Paris, France \\
$^{3}$ Center for Cosmology and Astro-Particle Physics, Ohio State University, Columbus, OH 43210\\
$^{4}$ Department of Physics, University of Illinois at Urbana-Champaign, Urbana, IL 61801\\
$^{5}$ Department of Physics \& Astronomy, University of Waterloo, Waterloo,
ON,  N2L 3G1 Canada \\
$^{6}$ Perimeter Institute for Theoretical Physics,
Waterloo, ON, N2L 2Y5, Canada \\
$^{7}$ Canadian Institute for Theoretical Astrophysics, 60 St. George St.,
Toronto, ON M5S 3H8 Canada \\
$^{8}$ Department of Astronomy, University of Illinois at Urbana-Champaign, Urbana, IL 61801\\
$^{9}$ Department of Astronomy, Ohio State University, Columbus, OH 43210\\
$^{10}$ Theoretical Division, Los Alamos National Laboratory, Los Alamos, NM 87545, USA
}}

\begin{document}

\maketitle

\label{firstpage}

\begin{abstract}
How do observed voids relate to  the underlying dark matter distribution?
To examine the spatial  distribution of dark matter contained within voids identified 
in galaxy surveys,
we apply Halo Occupation Distribution models representing sparsely and densely sampled galaxy surveys to a high-resolution $N$-body simulation.
We 
compare these galaxy voids to voids found in the halo distribution, 
low-resolution dark matter, and high-resolution dark matter.
We find that voids at all scales 
in densely sampled surveys --- and medium- to large-scale voids in 
sparse surveys ---
trace the same underdensities as dark matter, but they are 
larger in radius by $\sim 20\%$, 
they have somewhat shallower density profiles, and they have 
centers offset by $\sim 0.4 R_v$ rms.
However, in void-to-void comparison we find that shape estimators are less 
robust to sampling,
and the largest voids in sparsely sampled surveys suffer fragmentation 
at their edges.
We find that voids in galaxy surveys always correspond to  
underdensities in the dark matter, though the centers may be 
offset. When this offset is taken into account, we recover almost 
identical radial density profiles between galaxies and dark matter.
All mock catalogs used in this work are available 
at {http://www.cosmicvoids.net}.
\end{abstract}


\section{Introduction}

Current and future large-scale galaxy redshift 
survey programs map 
out large volumes 
of the cosmic web~\citep[e.g.,][]{Laureijs2011,Ahn2012, Parkinson2012}. 
Since galaxies trace 
the underlying dark matter distribution, albeit in a sparse 
and biased fashion~\citep{Baugh2013}, we can use statistics of the galaxy 
population (i.e., the overdensities) 
to understand the growth and present-day 
structure of the Universe and its relation to dark energy~\citep{Weinberg2012}.
Surveys typically focus on two-point statistics of the galaxy distribution, such as the 
galaxy autocorrelation function~\citep{Sanchez2012,Marin2013} or the 
baryon acoustic oscillation feature~\citep{Bassett2010}, both of which are 
sensitive probes of cosmology.

A complementary approach to using the structure of matter to 
understand cosmology is to examine the \emph{underdense}
regions of the cosmic web, namely cosmic voids~\citep{Gregory1978}. 
Voids offer 
several advantages over traditional measures of large-scale 
structure: they have a wide range of 
sizes~\citep{Hoyle2004, Pan2011, Sutter2012a}, 
allowing a study of multiple length scales simultaneously; 
they fill up most of the volume of the 
universe~\citep{Hoffman1982}, giving a large 
statistical weight; they are nearly empty of matter (by definition), 
so fifth forces from modified gravity and coupled dark matter-dark energy
remain unscreened inside them~\citep{Li2009,Li2012,Clampitt2013,Spolyar2013};
they can potentially serve as standard 
rulers~\citep{Hamaus2013}; and their 
statistical isotropy can be leveraged to perform an 
Alcock-Paczynski test~\citep{Alcock1979, Ryden1995, 
LavauxGuilhem2011, Sutter2012b}.

However, the statistical properties of voids in galaxy populations 
are not the same as those in dark matter 
distributions~\citep{Little1994,Benson2003,Sutter2013a}.
Since voids are defined by the lack of tracers (whether galaxies 
in a survey or particles in an $N$-body simulation), void 
sizes and shapes will necessarily be sensitive to the sampling 
density and biasing of the tracers. Additionally, voids exhibit a 
complex internal hierarchy of 
structure~\citep{Gottlober2003,Goldberg2004,Aragon2012} that is imperfectly sampled by galaxies.

While we can empirically 
model the changes to void statistics when transitioning 
from dark matter to galaxy 
voids (as we do in~\citealt{Sutter2013a}), 
we must ensure that galaxy 
surveys really are capable of robustly identifying voids. We must 
determine if a void traced by a galaxy survey of a given sampling density
corresponds to a single dark matter void. If it does, we must 
establish the relationship between the galaxy void and the 
dark matter underdensity. The nature of this relationship will 
depend on survey density, so we must also estimate the minimum 
galaxy density necessary to faithfully capture the portion of the 
cosmic web represented by voids.
This is necessary to interpret recent work such as~\citet{Melchior2013}, 
which probe the underlying dark matter potential with galaxy lensing.

There have been some previous efforts to examine the interiors 
of  voids, but these works focused on a single or 
very few voids at limited 
scales~\citep{Schmidt2001,Benson2003,Gottlober2003,Goldberg2004}. Most authors  attempt to 
make contact with observations by using mock galaxy populations within 
simulations~\citep{Benson2003,Ceccarelli2006,Pan2011,Tavasoli2013,Sutter2013c}.
Instead, in this work 
we examine in detail the interior dark matter contents of galaxy 
voids. We use a Halo Occupation Distribution model~\citep{Berlind2002}
to generate two mock galaxy surveys: a representative sparsely sampled survey
and a representative densely sampled survey. Drawing these mock surveys 
from a single high-resolution simulation, we compare on a void-by-void 
basis to voids in the halo distribution, low-resolution dark matter 
matched to the mean survey density, and high-resolution dark matter.

In the following section, we summarize our simulation, 
dark matter samples, mock galaxy catalogs, void finding technique, 
and method for matching galaxy voids to dark matter voids.
In Section~\ref{sec:interiors} we examine visually projected 
densities inside galaxy voids, trace the radial density profiles 
of dark matter within galaxy voids, and examine the correlations between 
underdensities in the galaxies and the dark matter at various radii.
Section~\ref{sec:matched} focuses on comparing galaxy voids and 
dark matter voids on an individual basis by examining relative 
positions, sizes, profiles, and ellipticities. Finally, 
in Section~\ref{sec:conclusions} we offer concluding comments and discussions
regarding prospects for upcoming galaxy redshift surveys.

\section{Numerical Approach}
\label{sec:approach}
\subsection{Simulations \& Mocks}
We source all samples and mock catalogs in this work from a single 
\lcdm~dark matter $N$-body simulation. 
We use the {\tt 2HOT} code, an adaptive treecode N-body method whose operation count
scales as $N \log N$ in the number of particles~\citep{warren13}.
Accuracy and error behavior have been improved significantly for
cosmological volumes through the use of a technique to subtract the
uniform background density, as well as using a compensating smoothing
kernel for small-scale force softening~\citep{dehnen01}.  We use a
standard symplectic integrator~\citep{quinn97} and an efficient
implementation of periodic boundary conditions using a high-order
($p=8$) multipole local expansion. We adjust the error tolerance
parameter to limit absolute errors to 0.1\% of the rms peculiar
acceleration. 
Initial conditions were
generated using a power spectrum calculated with {\tt CLASS}~\citep{blas11}
and realized with a modified version of {\tt 2LPTIC}~\citep{crocce06}.

This particular simulation assumed WMAP 7-year cosmological 
parameters~\citep{Komatsu2011}. The box size was 1~\gmpc~on a side 
and contained $1024^3$ particles, giving a particle mass resolution 
of $7.36 \times 10^{11} \hmsol$.
All analysis in this work used a single real-space snapshot at 
$z=0$.
For the dark matter analysis, we take successive random
subsamples of the particles to achieve tracer densities of
$10^{-2}$, $4 \times 10^{-3}$, and $3 \times 10^{-4}$ particles per 
cubic \hmpc. These samples as labeled as \emph{DM Full}, 
\emph{DM Dense}, and \emph{DM Sparse}, respectively.
We chose the latter values to roughly match the mean number densities 
of the mock galaxy populations, which we will discuss below.

We use the {\tt Rockstar} halo finder~\citep{behroozi13}, a 
six-dimensional phase-space plus
time halo finder, to identify spherical overdensity (SO) halos at 200
times the background density.  We use the default Rockstar parameters,
except for requiring strict SO masses that include unbound particles
and particles that may exist outside of the FOF group for the halo.
We use the halo catalog both as a direct source 
of tracers for void finding and as inputs for the HOD modeling.
We take two halo populations, one labeled \emph{Halos Dense} that
includes all halos down to the minimum resolvable halo mass of 
$1.47 \times 10^{12} \hmsol$ (20 particles), and one
labeled \emph{Halos Sparse}
that only includes halos above $1.2 \times 10^{13}~\hmsol$. 
We use these two thresholds to approximate the minimum mass used 
in the HOD distribution, thereby allowing us to compare voids found 
in halos to those found in galaxy populations.

We produce galaxy catalogs from the above halo population 
using the code described in~\citet{Tinker2006} and the HOD model 
described in~\citet{Zheng2007}. 
HOD modeling assigns central and satellite galaxies to a dark matter 
halo of mass $M$ according to a parametrized distribution.
In the case of the~\citet{Zheng2007} parametrization, the 
mean number of central galaxies is given by
\begin{equation}
\left\langle N_{\rm cen}(M)\right\rangle = \frac{1}{2} \left[
1 + \erf \left(\frac{\log M - \log M_{\rm min}}{\sigma_{\log M}}\right)
\right]
\end{equation}
and the mean number of satellites is given by
\begin{equation}
\left< N_{\rm sat}(M)\right> = \left\langle N_{\rm cen}(M) \right\rangle
\left( \frac{M-M_0}{M_1'}\right)^\alpha,
\end{equation}
where $M_{\rm min}$, $\sigma_{\log M}$, $M_0$, $M_1'$, and $\alpha$ 
are free parameters that can be inferred by fitting the observed space 
density and clustering of a given galaxy population.
The probability distribution of central galaxies $P(N_{\rm cen} | \left< 
N_{\rm cen} \right>)$ is a nearest-integer 
distribution, and satellites follow a Poisson
$P(N_{\rm sat} | \left< N_{\rm sat} \right>)$.

Using the above model we generate two mock catalogs.
The first is matched to the 
SDSS DR9 CMASS galaxy sample~\citep{Ahn2012} using the parameters 
found by~\citet{Manera2013} ($\sigma_{\log M} = 0.596$, $M_0 = 1.2\times10^{13}\hmsol$, $M_1' = 10^{14}\hmsol$, $\alpha = 1.0127$, and $M_{\rm min}$ chosen to fit the mean number density). We call this sample \emph{HOD Sparse}, 
since we are using it to represent a relatively low-resolution 
galaxy sample.
We use the full resolved halo population to create this sample, which 
contains many galaxies in halos with $M < M_0$ because of the 
large scatter parameter $\sigma_{\log M}$.
Our second catalog, named \emph{HOD Dense}, 
is matched to the SDSS DR7 main sample~\citep{Strauss2002} at $z<0.1$
using the parameters found by~\citet{Zehavi2011} for galaxies with 
$M_r < -21 + 5 \log h$ 
($\sigma_{\log M} = 0.21$, $M_0 = 6.7\times10^{11}\hmsol$, 
$M_1' = 2.8 \times 10^{13}\hmsol$, $\alpha = 1.12$). 
Our simulation only resolves halos down to $M \approx 2 M_0$, so 
we cannot fully represent the HOD, but adopting these parameters 
and a cutoff at the 20-particle minimum halo mass produces a high density
galaxy catalog with reasonably realistic clustering properties.

\subsection{Void Finding}
We identify voids with a modified version of 
{\tt ZOBOV}~\citep{Neyrinck2008, LavauxGuilhem2011, Sutter2012a}.
These modifications include handling of survey masks, performance 
enhancements, enforcement of bijectivity 
 in the construction of the Voronoi graph 
in high-density regimes, and the development of a fully-pipelined 
environment that handles input preparation and post-processing 
filtering. 

{\tt ZOBOV} creates a Voronoi tessellation of the tracer particle 
population and uses the watershed transform to group Voronoi 
cells into zones and voids~\citep{Platen2007}. 
The watershed transform identifies catchment basins 
as the cores of voids, and ridgelines, which separate the flow 
of water, as the boundaries of voids.
The watershed transform naturally builds a nested hierarchy of 
voids~\citep{LavauxGuilhem2011, Bos2012}, and for the purposes 
of this work we only examine \emph{root} voids, which 
are voids at the base of the tree hierarchy and hence have no parents.
We also impose two density-based criteria on our void catalog. 
The first is a threshold cut within {\tt ZOBOV} itself where
voids only include as additional members Voronoi zones with density
less than $0.2$ the mean particle density.
If a void consists of only a single zone (as they often do in
sparse populations) then this restriction does not apply.
We apply the second density criterion as a post-processing step:
we only include voids with mean central densities below $0.2$ times
the mean particle density. We measure this central density 
within a sphere of radius $R = R_{\rm eff}/4$, where
\begin{equation}
  R_{\rm eff} \equiv \left( \frac{3}{4 \pi} V \right)^{1/3},
\end {equation}
where $V$ is the total volume of the Voronoi cells 
that contribute to the void.
We also ignore voids with $R_{\rm eff}$ below the mean particle 
spacing $\bar{n}^{-1/3}$ of the tracer population, as these will arise simply 
from Poisson fluctuations. 
In sum, we identify voids as depressions in the tracer density
(of dark matter particles, halos, or galaxies); voids are 
non-spherical aggregations of Voronoi cells that share a common 
basin and are bounded by a common set of higher-density walls.

Additionally, for the analysis below we need to define a center for the 
void. For our work, we take the barycenter, or volume-weighted 
center of all the Voronoi cells in the void:
\begin{equation}
  {\bf X}_v = \frac{1}{\sum_i V_i} \sum_i {\bf x}_i V_i,
\label{eq:barycenter}
\end{equation}
where ${\bf x}_i$ and $V_i$ are the positions and Voronoi volumes of 
each tracer $i$, respectively.

Table~\ref{tab:voidsamples} summarizes the samples used in this work, 
their minimum effective void radius ($R_{\rm eff, min} \equiv \bar{n}^{-1/3}$), 
and the number of voids identified in the simulation 
volume.

\begin{table}
\centering
\caption{Summary of sample void populations.}
\tabcolsep=0.11cm
\footnotesize
\begin{tabular}{ccccc}
  Dataset Type & Sample Name & $R_{{\rm eff}, {\rm min}}$ (\hmpc) & $N_{\rm voids}$ &\\
  \hline  \hline
DM &  DM Full & 5 & 42948 & \\
\hline
DM &  DM Dense & 7 & 21865 & \\
Halos &  Halos Dense & 7 & 11419 & \\
HOD &  HOD Dense & 7 & 9503 & \\
\hline
DM &  DM Sparse & 14 & 2611 & \\
Halos &  Halos Sparse & 14 & 2073 & \\
HOD &  HOD Sparse & 14 & 1422 & \\
\hline
\end{tabular}
\label{tab:voidsamples}
\end{table}


\subsection{Void Matching}
\label{subsec:matching}

We use the amount of volume overlap to find matches between galaxy 
voids and voids in other samples. For each galaxy void, we first 
make a list of potential matches by considering all voids in the 
matched void catalog whose centers lie within the watershed volume of the 
galaxy void. Then for each potential matched void, we sum the volumes 
of each particle whose Voronoi cell overlaps any Voronoi cell of the 
galaxy void. We take the potential matched void with the greatest amount of 
volume overlap as the match. 

To simplify the measurement of 
overlap between two Voronoi cells, we place each particle at the center 
of a sphere whose volume is the same as its respective cell. We 
measure the distance between particles and assume they overlap if 
their distances meet the criterion
\begin{equation}
  d \le 0.25 ( R_1 + R_2 ),
\end{equation}
where $d$ is the distance and $R_1$ and $R_2$ are the radii of the spheres 
assigned to particles 1 and 2, respectively. This procedure is 
approximate, but adequate for our purposes in this paper. 
We found the factor 
of $0.25$ to strike the best balance between conservatively estimating 
overlap while still accounting for our rough estimate 
of the Voronoi volume of each particle.

To account for the possibility that dark matter voids may be less underdense 
(in terms of central density) than galaxy voids, we allow matching between 
galaxy voids and any void in the matched catalog, including 
voids that would normally by removed by the central density 
threshold discussed in the previous section.

\section{Void interiors}
\label{sec:interiors}

\begin{figure*} 
  \centering 
   {\includegraphics[type=png,ext=.png,read=.png,width=0.48\textwidth]{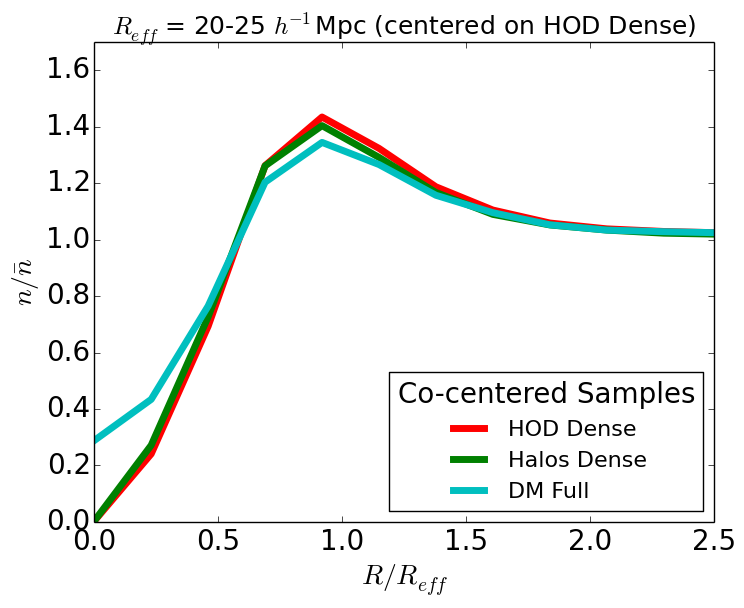}}
  {\includegraphics[type=png,ext=.png,read=.png,width=0.48\textwidth]{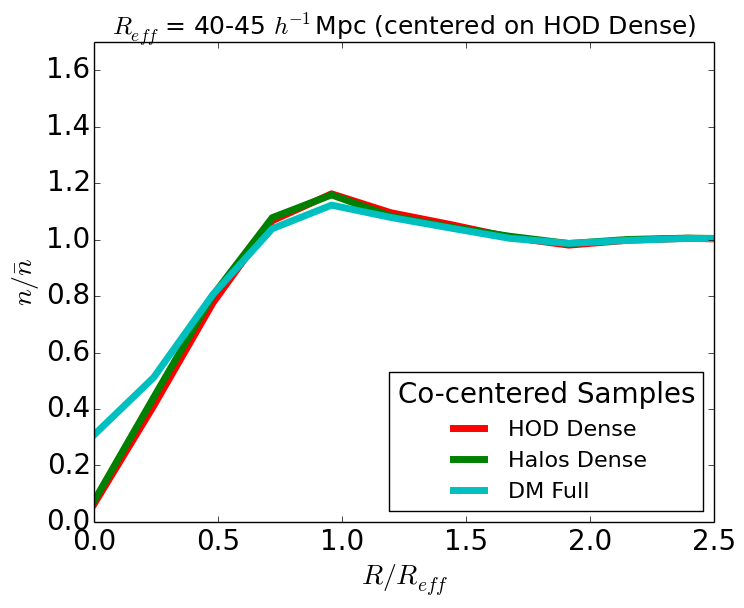}}
  {\includegraphics[type=png,ext=.png,read=.png,width=0.48\textwidth]{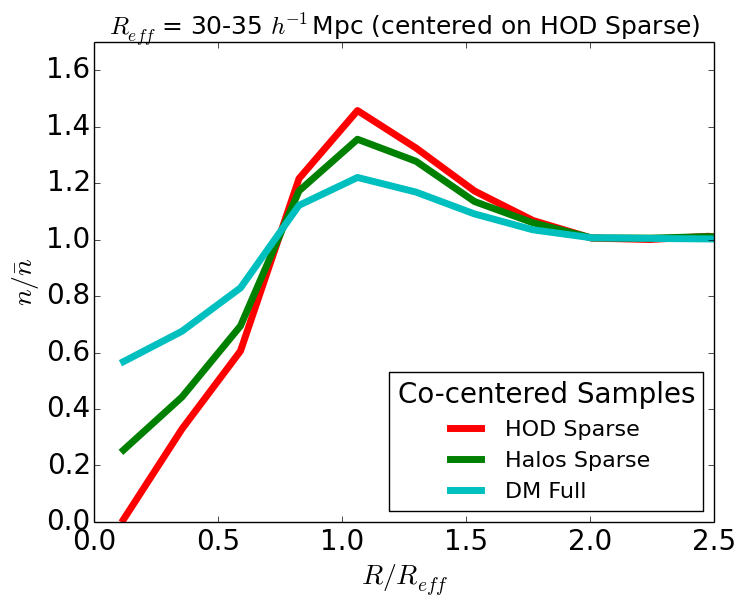}}
  {\includegraphics[type=png,ext=.png,read=.png,width=0.48\textwidth]{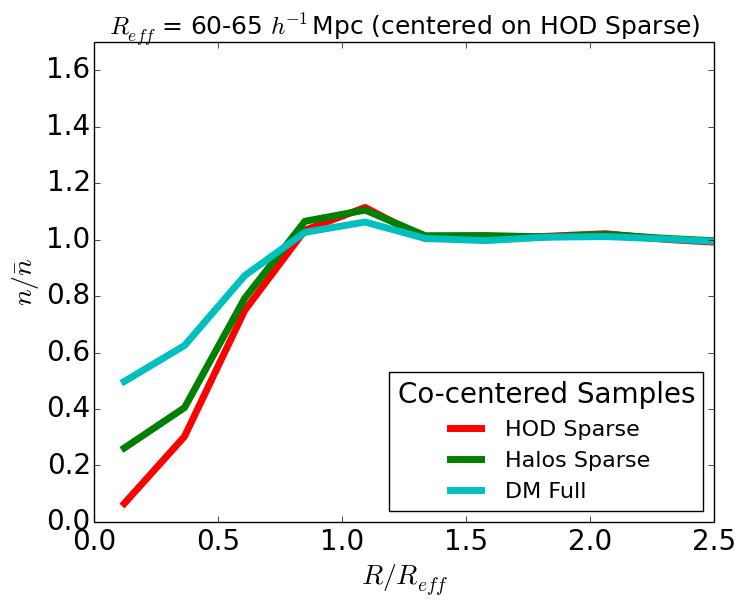}}
 \caption{Radial density profiles of stacked 
           voids in the \emph{HOD Dense} (top 
           row) and \emph{HOD Sparse} (bottom row) samples. For the 
           other samples listed, we keep the centers fixed on the voids 
           identified in the respective \emph{HOD} samples. 
           We choose stacks to highlight overcompensated (left column) 
           and undercompensated (right column) void scales,
           and we indicate the void size ranges used in each stack at the 
           top of each plot. 
           We normalize the profile in each sample to the
           mean number density $\bar{n}$ of that sample.
           $R_{\rm eff}$ corresponds to the median
           void size in the stack (i.e., there is no rescaling of the voids).
          }
\label{fig:1dprofile_cocenter}
\end{figure*}

In Figure~\ref{fig:1dprofile_cocenter} we 
compare the radial density profiles of voids identified in the HOD mock 
galaxy samples to the radial profiles of halos and dark matter 
about the same centers.
To stack voids we align the 
barycenters of all voids in a given size range.
We keep the void positions fixed as we build profiles in the 
halo and dark matter samples. For the dark matter we only show 
the \emph{DM Full} sample, as the sparsely sampled dark matter would 
necessarily yield the same average result but with greater noise.

Beginning with the \emph{HOD Dense} voids (top row), we see that the 
radial density profiles of HOD galaxies and halos are virtually 
identical. In the smallest radius bin ($R_{\rm eff} = 20-25$ \hmpc) 
these spherically averaged profiles rise from $n/\bar{n} \approx 0$ 
at $R=0$ to a maximum of $n/\bar{n} \approx 1.4$ at $R=0.9 R_{\rm eff}$, 
with a clear compensation shell surrounding the central underdensity.
For the larger size sample the compensation is much weaker and 
the slope is shallower.
The central dark matter densities are $n/\bar{n} \approx 0.3$ 
for all void radii, indicating that a combination of galaxy bias and shot 
noise has allowed the watershed approach to identify regions 
that are more underdense in galaxies than in dark matter. 
However, the dark matter profiles match the galaxy and halo 
profiles past $R \approx 0.4 R_{\rm eff}$. Importantly, the 
voids that are large in galaxies are also large in dark matter.

Turning to the \emph{HOD Sparse} sample (bottom row), the galaxy 
central densities are now lower than the halo central densities, 
which suggests that the watershed algorithm is taking advantage of 
sampling fluctuations to find deep minima in the galaxy distribution.
The central dark matter densities are higher than in the 
\emph{Dense} case ($n/\bar{n} \approx 0.5$), and the dark matter profiles 
are shallower than the galaxy profiles. The compensation effect is also 
weaker. Not surprisingly, voids in this much sparser, 
more highly biased galaxy population are less effective (relative to 
\emph{HOD Dense}) at corresponding to deep minima in the 
dark matter distribution. Nonetheless, the voids in the galaxy 
distribution are indeed underdense in the halos and dark matter, 
with similar radial extents, demonstrating the ability of a 
BOSS-like galaxy survey to reveal large-scale matter underdensities. 

\begin{figure*} 
  \centering 
  {\includegraphics[type=png,ext=.png,read=.png,width=0.32\textwidth]{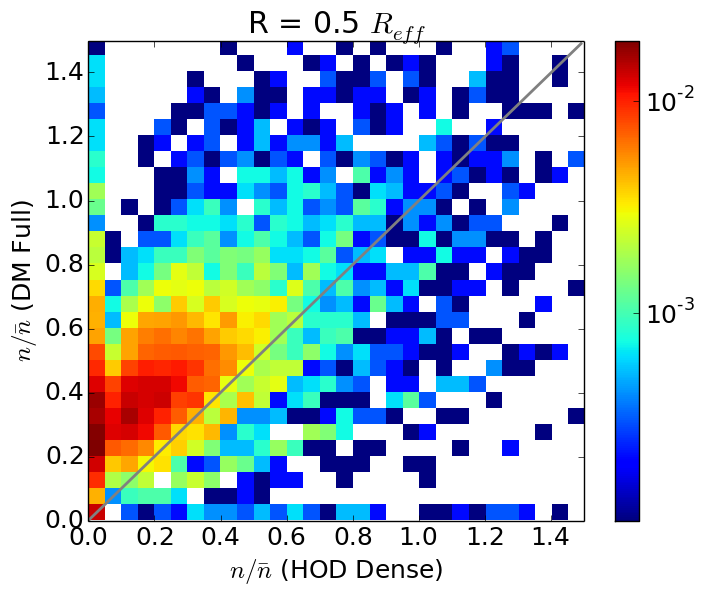}}
  {\includegraphics[type=png,ext=.png,read=.png,width=0.32\textwidth]{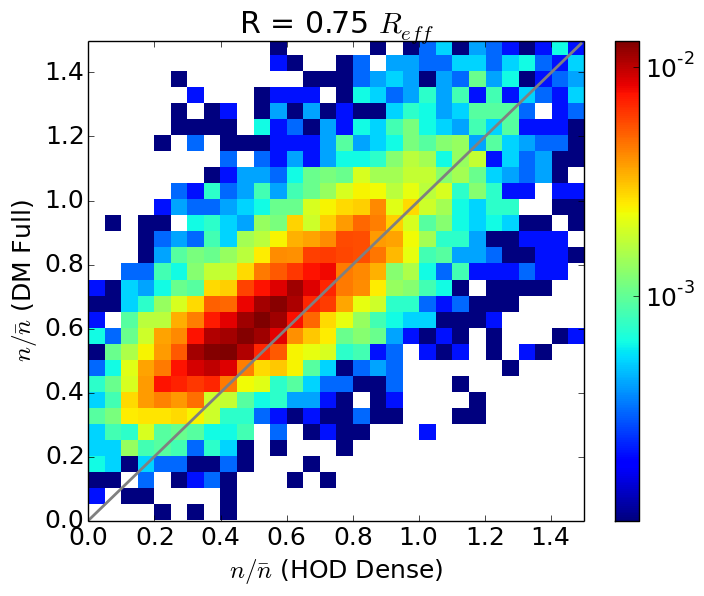}}
  {\includegraphics[type=png,ext=.png,read=.png,width=0.32\textwidth]{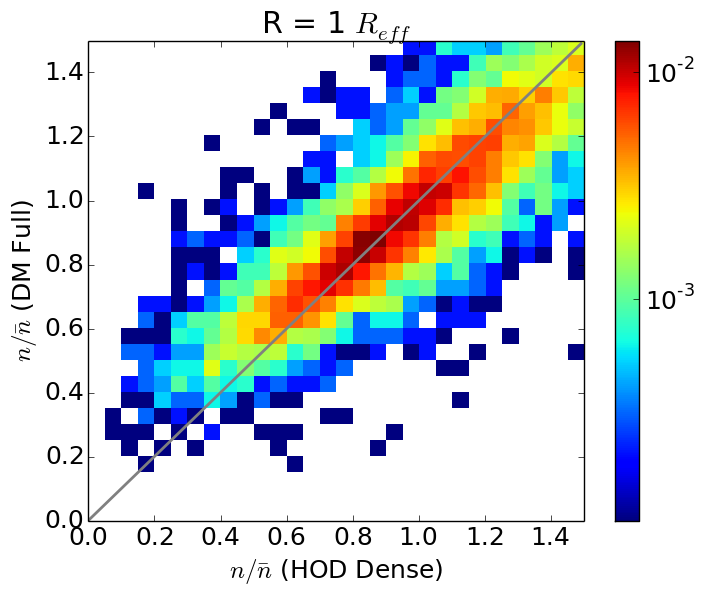}}
  {\includegraphics[type=png,ext=.png,read=.png,width=0.32\textwidth]{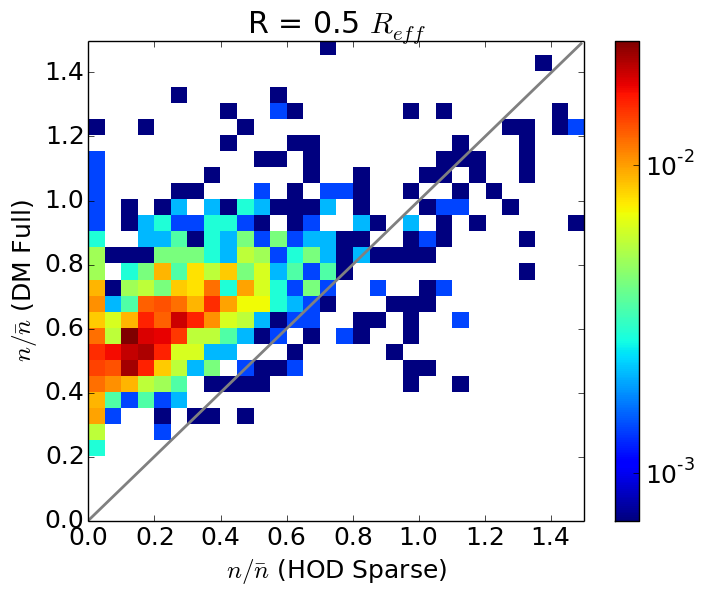}}
  {\includegraphics[type=png,ext=.png,read=.png,width=0.32\textwidth]{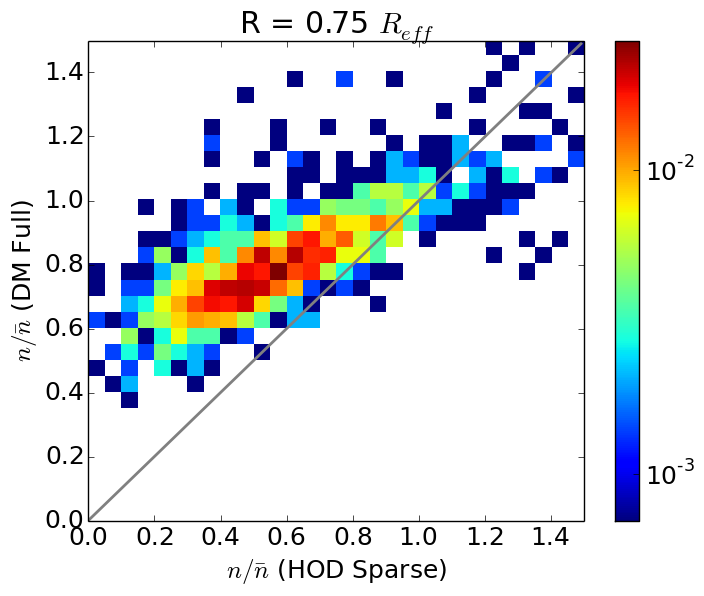}}
  {\includegraphics[type=png,ext=.png,read=.png,width=0.32\textwidth]{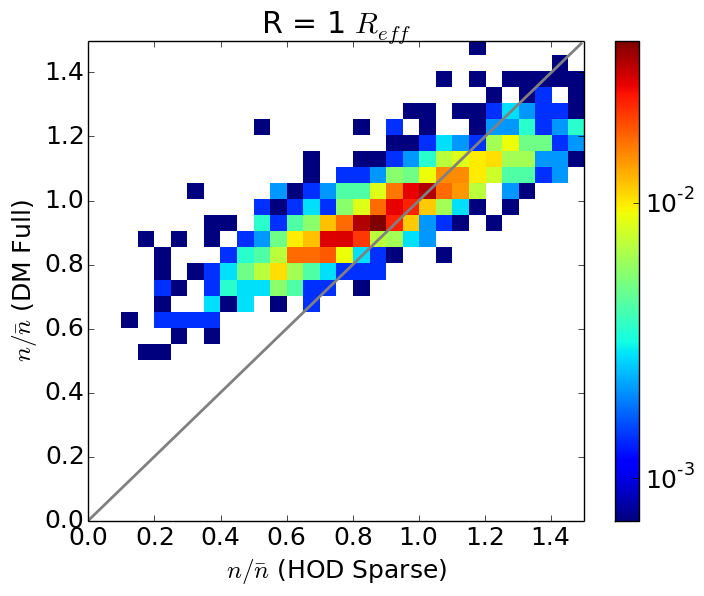}}
  \caption{Correlations between dark matter and galaxy densities with 
           various radii for the 
           \emph{HOD Dense} (top row) and
           \emph{HOD Sparse} (bottom row) void samples.
           We compute mean densities in spheres of radius
           $0.5 R_{\rm eff}$ (left),
           $0.75 R_{\rm eff}$ (middle), and
           $1.0 R_{\rm eff}$ (right).
           Densities in each sample are relative to the mean
           number density $\bar{n}$ of that sample.
           Each histogram bin has width $0.05 n/\bar{n}$.
           Correlation values are the fraction of all 
           voids in each bin, and the 
           thin grey line indicating equal densities is to guide the eye.
            }
\label{fig:underdencorr}
\end{figure*}

To further quantify this relationship,
in Figure~\ref{fig:underdencorr} 
we correlate densities measured within various radii in the 
galaxy voids to densities measured in the high-resolution 
dark matter. That is, we compute $n/\bar{n} (< R/R_{\rm eff})$, 
spherically averaged, for each galaxy void. We then compute 
$n/\bar{n}$ in the \emph{DM Full} sample within those 
same spheres, and plot the correlations between the pairs 
of densities. 
We measure the densities at $0.5$, 
$0.75$, and $1.0$ times the effective radius $R_{\rm eff}$ of each 
galaxy void, considering all voids above the minimum $R_{\rm eff}$ 
threshold.

In Figure~\ref{fig:underdencorr} 
we see that even the most underdense cores of \emph{HOD Dense} 
galaxy voids 
correlate with higher densities ($\sim 0.3~n/\bar{n}$) in the dark matter.
Within $R=0.75 R_{\rm eff}$, a strong correlation appears, and the 
densities in the galaxies and the dark matter are nearly identical, 
modulated by a slight offset. Within the full void effective radius, 
however, the galaxy and dark matter densities match excellently.
At this radius, a density  
in the galaxies corresponds to an equivalent density in the 
dark matter but with $\sim 10\%$ rms scatter.

The correlations to \emph{HOD Sparse} galaxy voids show similar trends, 
but with larger systematic offsets, even at the void effective 
radius.
Overall, voids with low densities at $R=0.5$-$1.0 R_{\rm eff}$ 
correlate most strongly with high-resolution dark matter 
densities of 0.5-0.8 $n/\bar{n}$. 

\begin{figure*} 
  \centering 
  {\includegraphics[type=png,ext=.png,read=.png,width=0.24\textwidth]{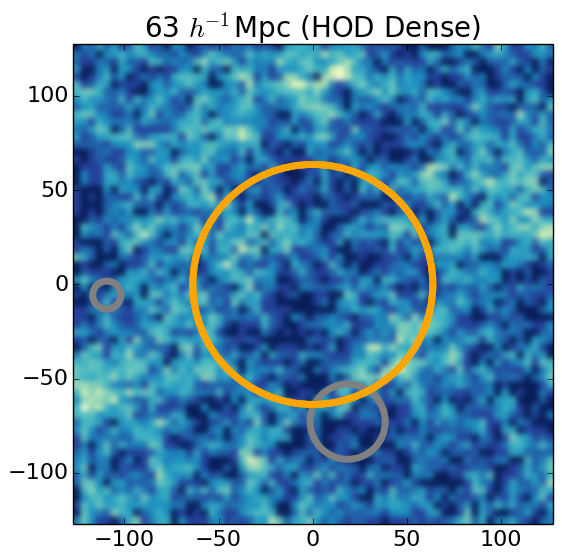}}
  {\includegraphics[type=png,ext=.png,read=.png,width=0.24\textwidth]{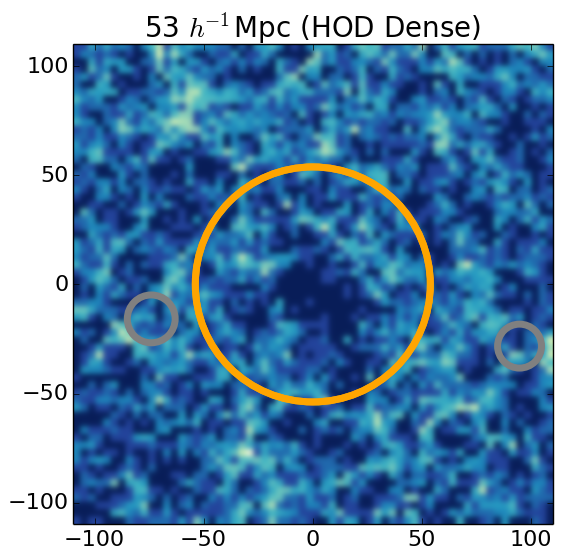}}
  {\includegraphics[type=png,ext=.png,read=.png,width=0.24\textwidth]{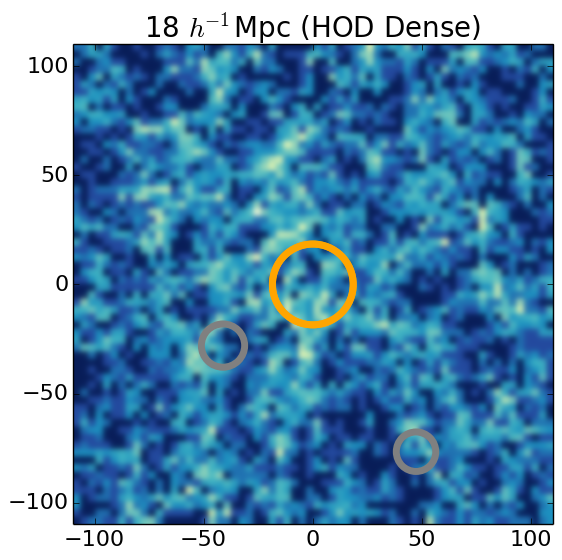}}
  {\includegraphics[type=png,ext=.png,read=.png,width=0.24\textwidth]{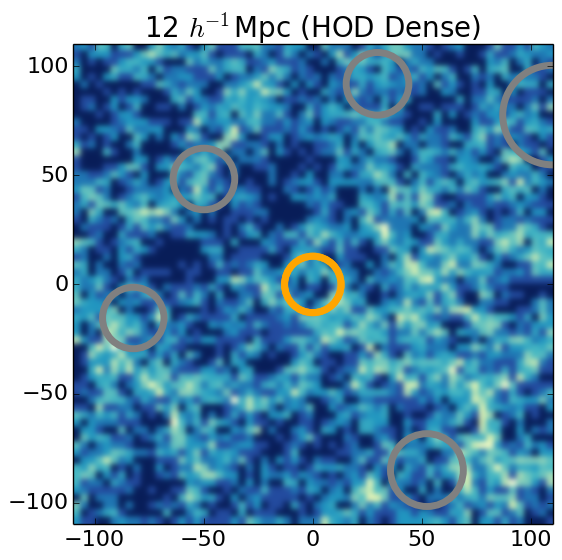}}
  {\includegraphics[type=png,ext=.png,read=.png,width=0.24\textwidth]{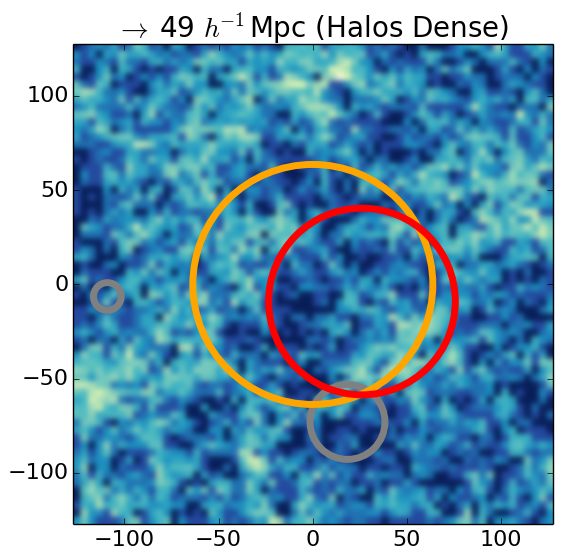}}
  {\includegraphics[type=png,ext=.png,read=.png,width=0.24\textwidth]{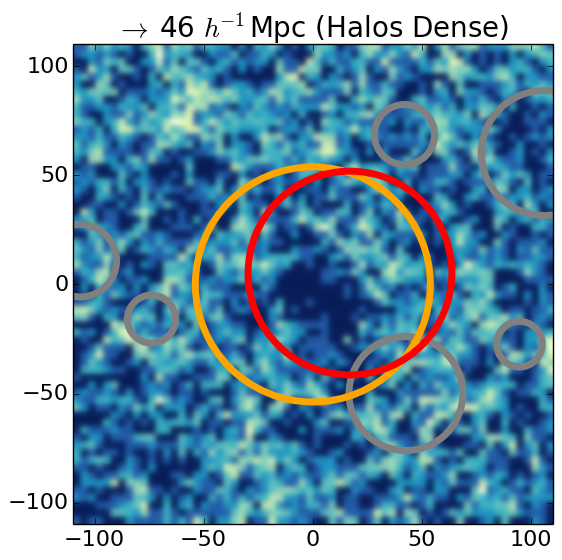}}
  {\includegraphics[type=png,ext=.png,read=.png,width=0.24\textwidth]{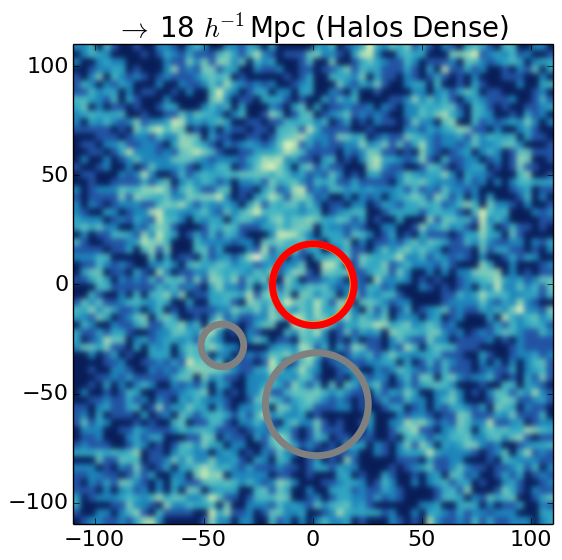}}
  {\includegraphics[type=png,ext=.png,read=.png,width=0.24\textwidth]{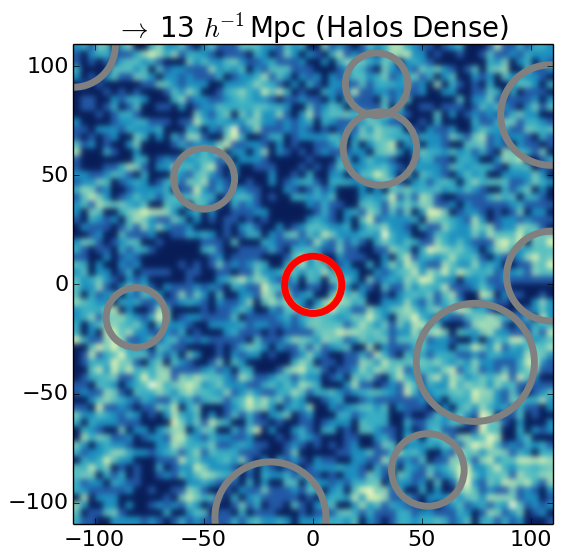}}
  {\includegraphics[type=png,ext=.png,read=.png,width=0.24\textwidth]{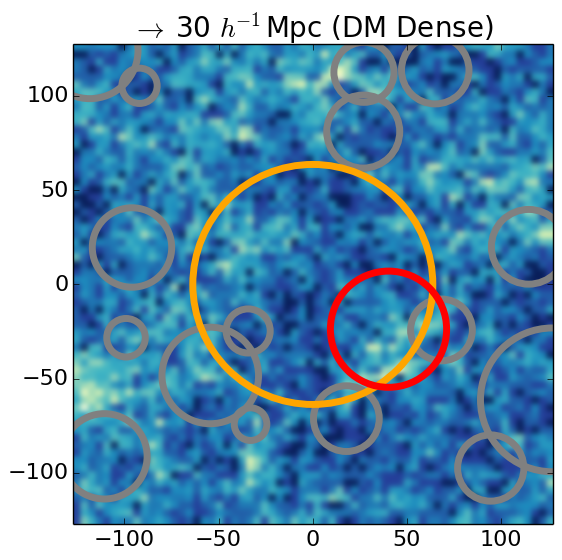}}
  {\includegraphics[type=png,ext=.png,read=.png,width=0.24\textwidth]{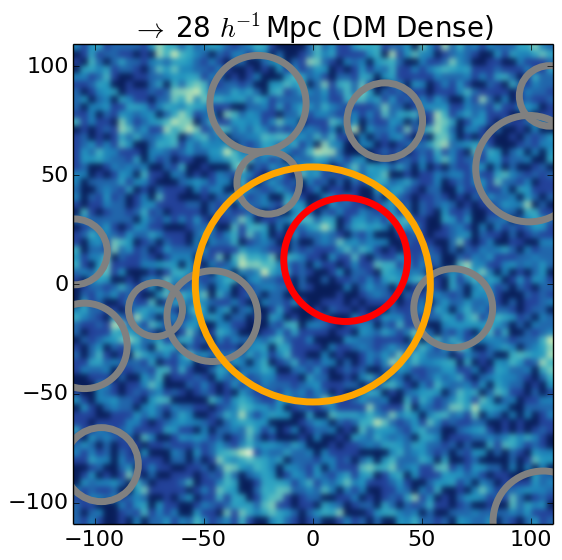}}
  {\includegraphics[type=png,ext=.png,read=.png,width=0.24\textwidth]{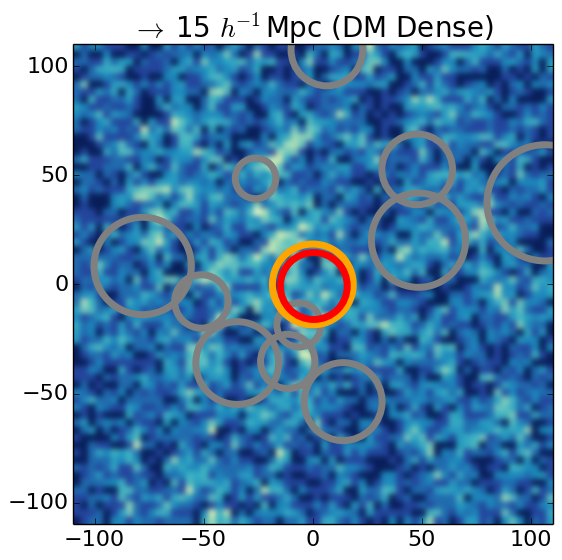}}
  {\includegraphics[type=png,ext=.png,read=.png,width=0.24\textwidth]{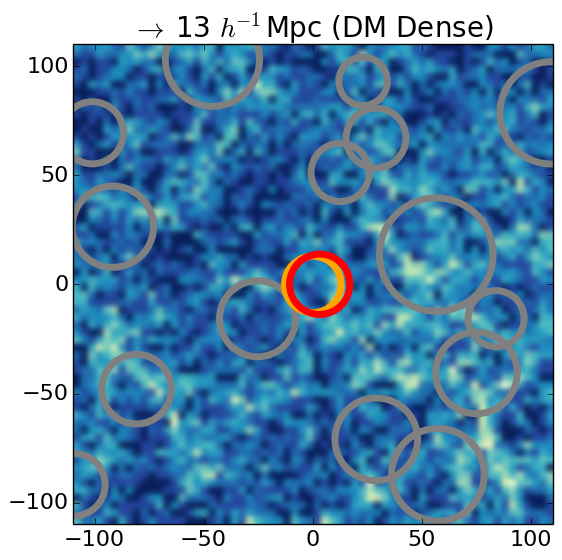}}
  {\includegraphics[type=png,ext=.png,read=.png,width=0.24\textwidth]{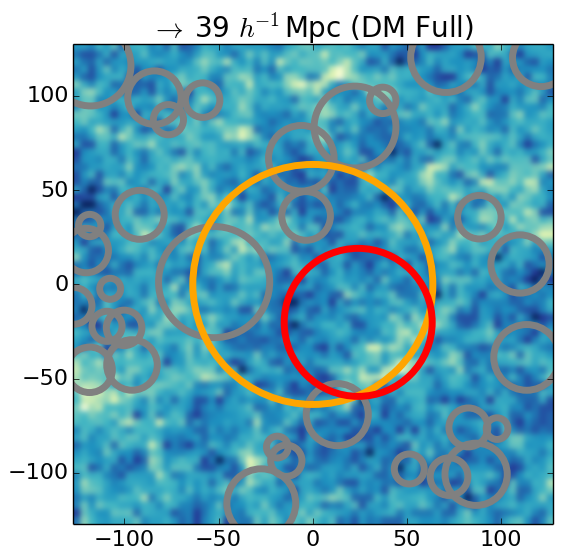}}
  {\includegraphics[type=png,ext=.png,read=.png,width=0.24\textwidth]{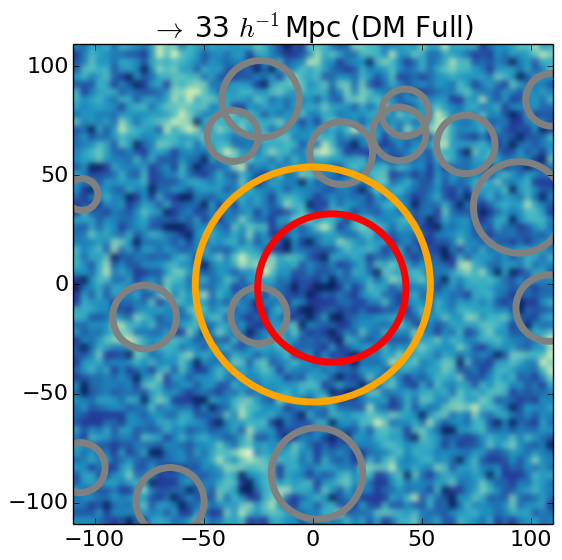}}
  {\includegraphics[type=png,ext=.png,read=.png,width=0.24\textwidth]{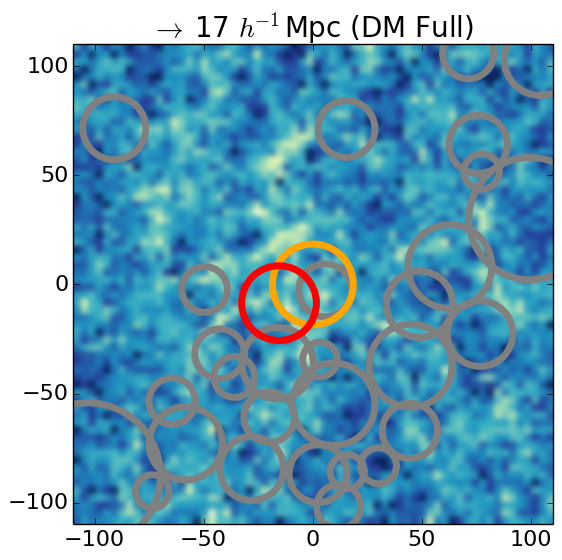}}
  {\includegraphics[type=png,ext=.png,read=.png,width=0.24\textwidth]{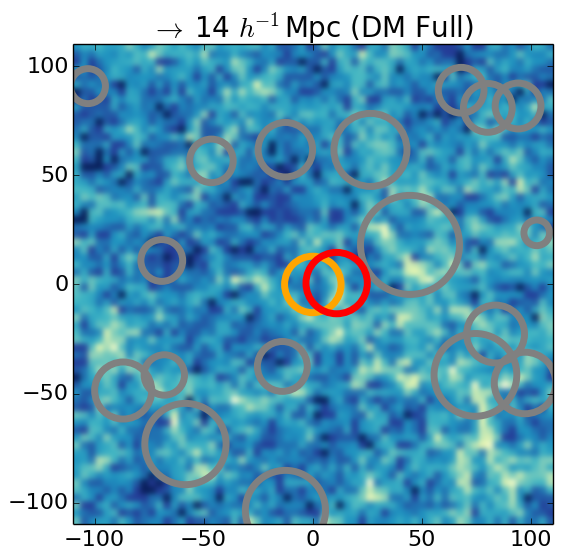}}
  \caption{Voids in the \emph{HOD Dense} sample and 
           projected density slices of various tracer populations.
           We select four voids at random but chosen to fairly 
           represent the size ranges in this sample.
           We list the size of each void in the top row.
           We represent the selected void as an orange circle with 
           radius equal to its effective radius $R_{\rm eff}$.
           Other unrelated voids in the slice are plotted as grey circles.
           We project each \emph{HOD Dense} void on other tracer populations 
           in the next three rows. The second row is the 
           \emph{Halos Dense} sample, the third is 
           \emph{DM Dense}, and the bottom row is \emph{DM Full}.
           Within each sample, the void identified as the best match
           using the procedure discussed in Section~\ref{subsec:matching}
           is plotted in red.
           The size of 
           the best-match void and the name of the sample is given 
           on the top of each plot.
           Note that in some cases there is no match found, which 
           we indicate by ``No match'', 
           and in others there is a perfect correspondence,
           so that the orange circle is not visible.
           The width of each slice along the line of sight is 
           50~\hmpc.
           The tracer
           densities are given in units of $\bar{n}$, the mean number 
           density of each sample, and colored from 0 (dark blue) to 
           $1.5$ (white). The axes are marked in units of \hmpc.
         }
\label{fig:denmap_highres}
\end{figure*}

\begin{figure*} 
  \centering 
  {\includegraphics[type=png,ext=.png,read=.png,width=0.24\textwidth]{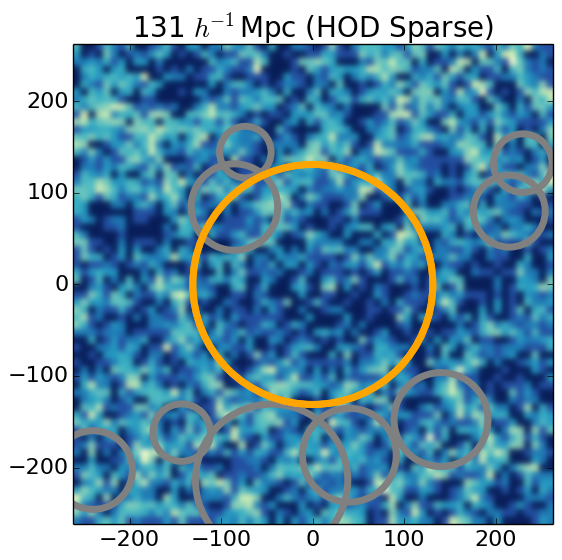}}
  {\includegraphics[type=png,ext=.png,read=.png,width=0.24\textwidth]{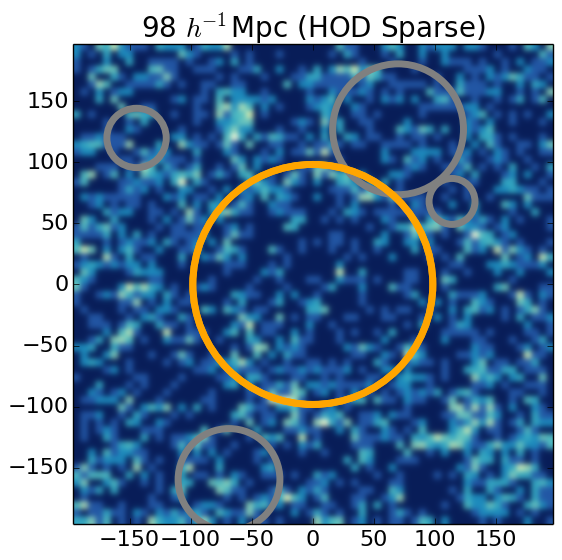}}
  {\includegraphics[type=png,ext=.png,read=.png,width=0.24\textwidth]{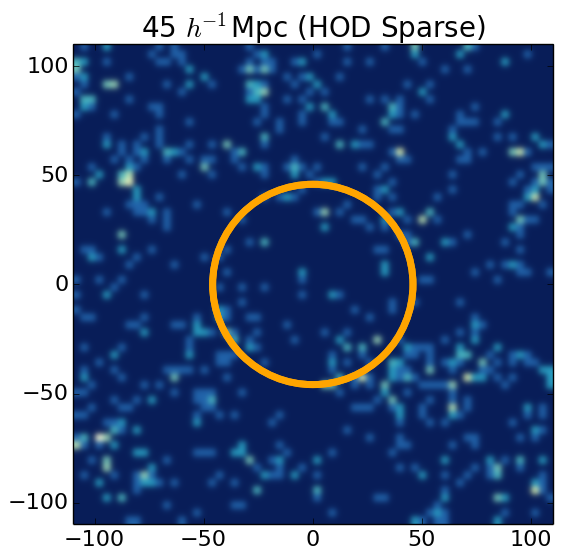}}
  {\includegraphics[type=png,ext=.png,read=.png,width=0.24\textwidth]{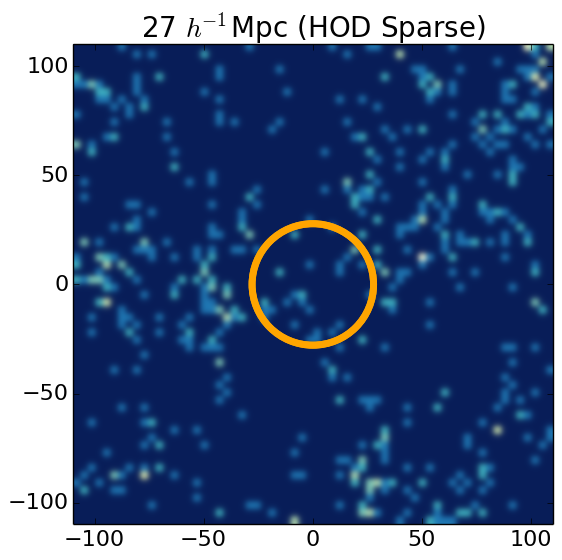}}
  {\includegraphics[type=png,ext=.png,read=.png,width=0.24\textwidth]{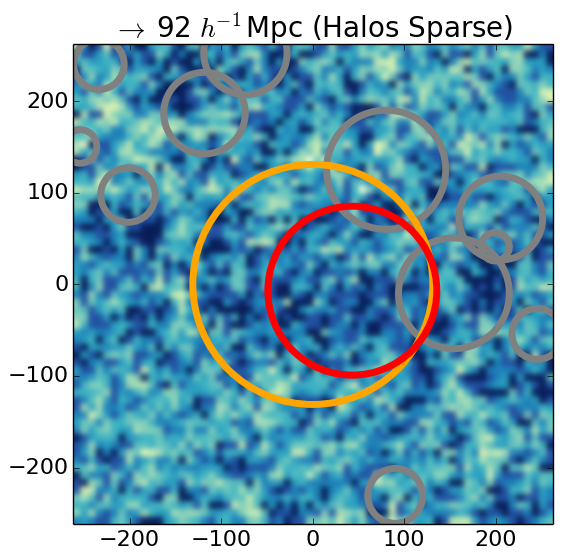}}
  {\includegraphics[type=png,ext=.png,read=.png,width=0.24\textwidth]{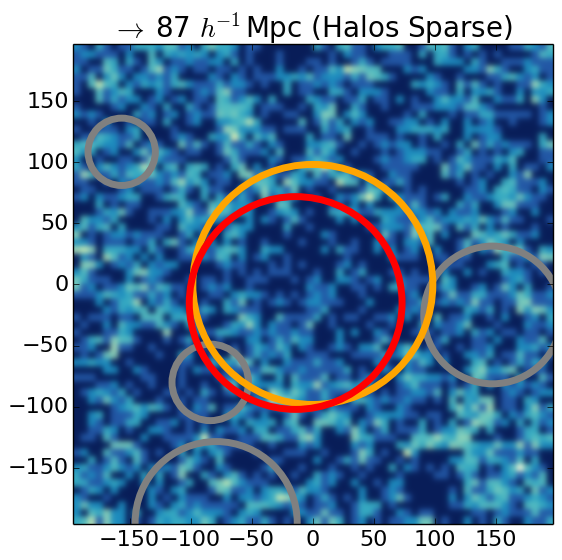}}
  {\includegraphics[type=png,ext=.png,read=.png,width=0.24\textwidth]{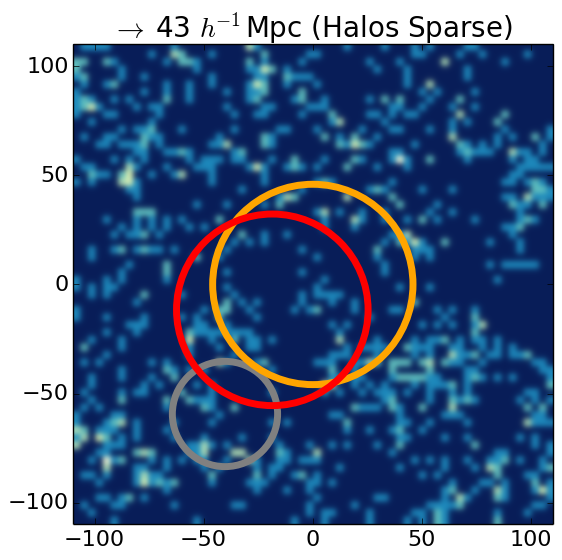}}
  {\includegraphics[type=png,ext=.png,read=.png,width=0.24\textwidth]{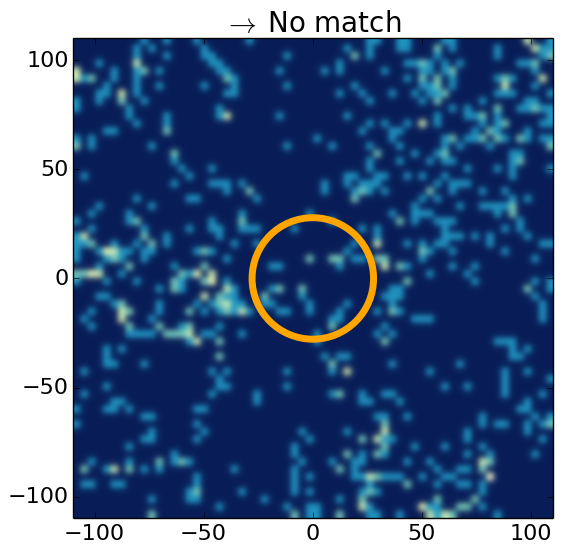}}
  {\includegraphics[type=png,ext=.png,read=.png,width=0.24\textwidth]{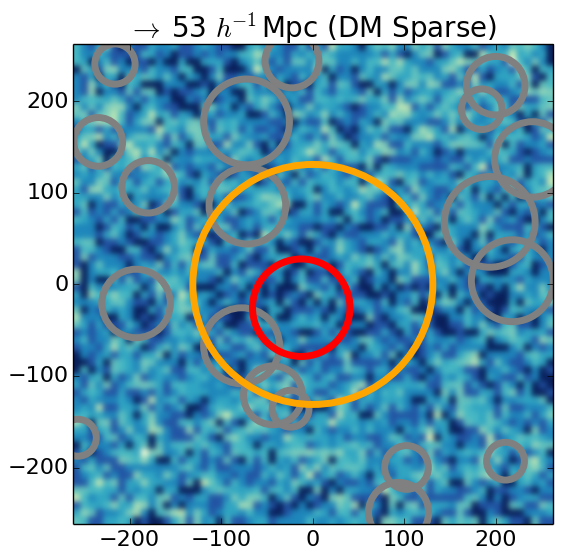}}
  {\includegraphics[type=png,ext=.png,read=.png,width=0.24\textwidth]{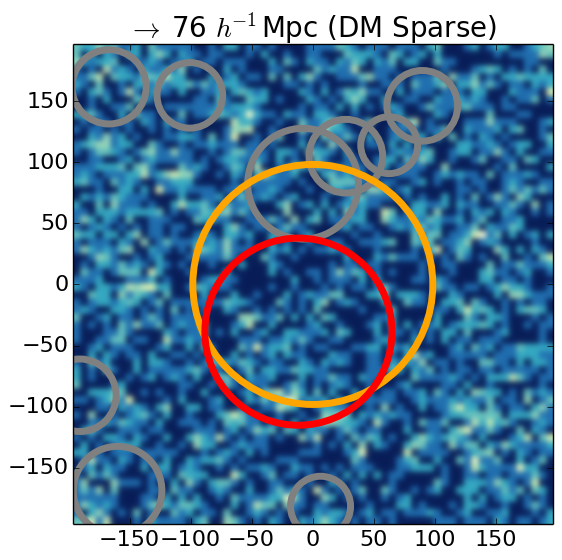}}
  {\includegraphics[type=png,ext=.png,read=.png,width=0.24\textwidth]{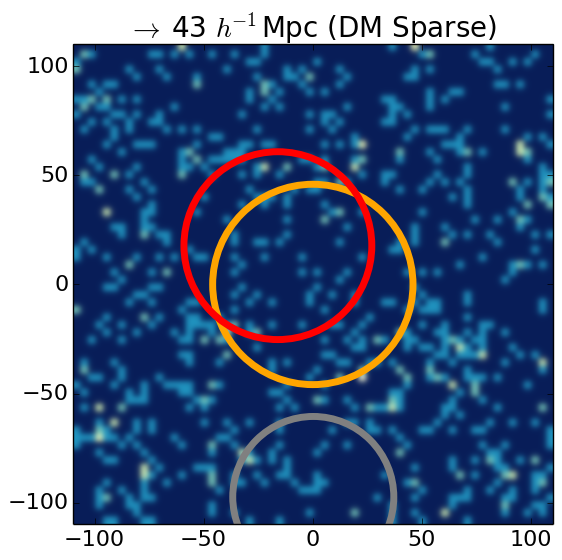}}
  {\includegraphics[type=png,ext=.png,read=.png,width=0.24\textwidth]{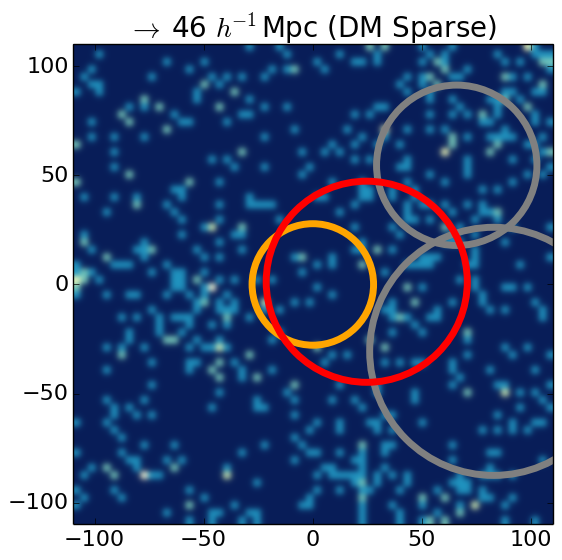}}
  {\includegraphics[type=png,ext=.png,read=.png,width=0.24\textwidth]{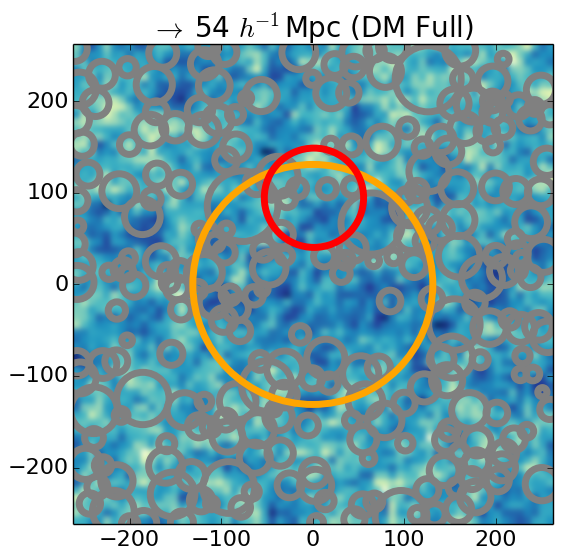}}
  {\includegraphics[type=png,ext=.png,read=.png,width=0.24\textwidth]{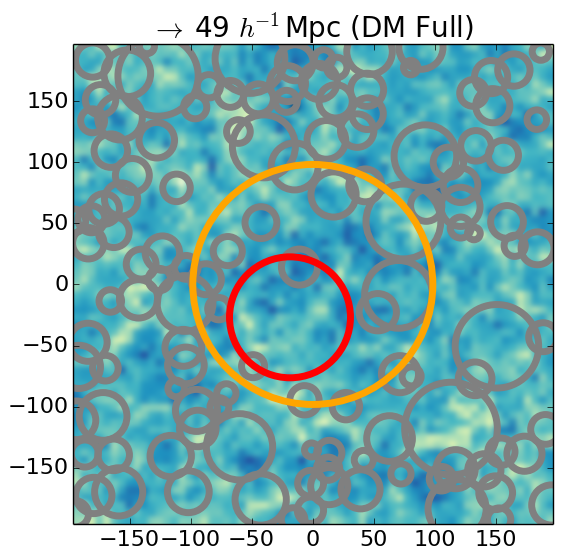}}
  {\includegraphics[type=png,ext=.png,read=.png,width=0.24\textwidth]{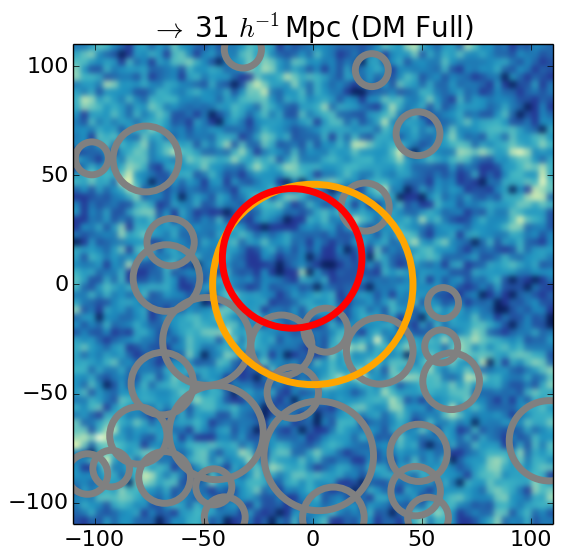}}
  {\includegraphics[type=png,ext=.png,read=.png,width=0.24\textwidth]{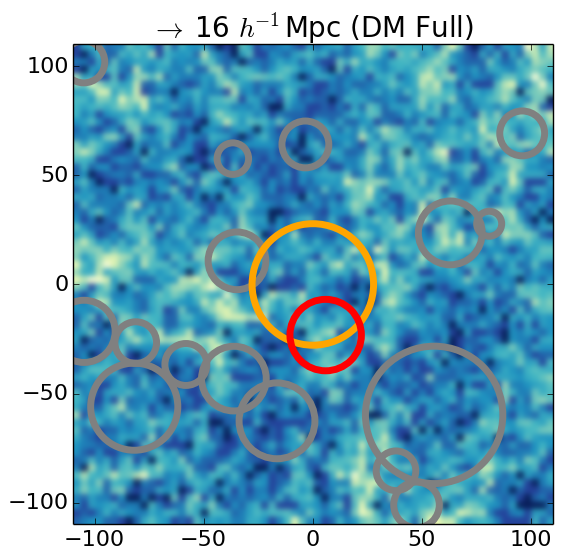}}
  \caption{Same as Figure~\ref{fig:denmap_highres} but for the 
           \emph{HOD Sparse} sample.
          }
\label{fig:denmap_lowres}
\end{figure*}

We finally turn to slices of the tracer density maps to give a visual 
impression of the behavior of the dark matter inside galaxy voids.
Figures~\ref{fig:denmap_highres} and~\ref{fig:denmap_lowres}
show such slices, which we have centered on four voids
from the \emph{HOD Dense} and \emph{HOD Sparse} samples.
We chose the particular voids randomly 
but selected a representative sample from 
the range of scales in the void catalog.
We overplot each void on its host sample density map, as well as 
density maps from the halo catalog corresponding to the mock 
galaxy sample (\emph{Halos Dense} for \emph{HOD Dense} and 
\emph{Halos Sparse} for \emph{HOD Sparse}), 
the low-resolution dark matter sample with an equivalent 
mean number density (\emph{DM Dense} for \emph{HOD Dense} and 
\emph{DM Sparse} for \emph{HOD Sparse}), 
and the high-resolution dark matter 
sample \emph{DM Full}.
We also show any nearby voids that happen to lie in the slice, and 
the void chosen to be the best match using the procedure described 
in the section above. We represent each void as a circle with 
radius $R_{\rm eff}$.

While the plotted circles only crudely represent the complex shape of each 
void (see, for example, Figure 2 of~\citealt{Sutter2012a}), they do 
give us a useful base for examining void contents. 

The voids traced by the \emph{HOD Dense} sample tend 
to remain in the dark matter at all void scales, 
as we see in Figure~\ref{fig:denmap_highres}.
While there is modest fragmentation (i.e., the breakup of a single 
galaxy void into many smaller dark matter substructures) in the 
dark matter, it is much less severe than in the \emph{HOD Sparse}
case and does not strongly affect the void interiors. 
Matched voids in the halo, low-resolution dark matter, and 
high-resolution dark matter 
are smaller, but still correspond 
to the same fundamental structure identified in the galaxies.
For the smallest voids there is an almost perfect match between 
the galaxy voids and their dark matter counterparts.
We see the surrounding and 
internal structures thicken in the dark matter, but since 
the galaxies in this case already faithfully represent the cosmic 
web, it does not lead to large distortions of the voids.

In the case of the \emph{HOD LowRes} sample shown in 
Figure~\ref{fig:denmap_lowres}, we see that while large 
voids are situated in clear underdense regions, the galaxies are so sparse
that it is difficult to visually separate the void and wall regions.
The largest voids begin to fragment into smaller structures 
even in the halo and low-resolution dark matter populations, 
while we have difficulty finding appropriate matches for the smallest 
voids. The walls surrounding the underdense regions marked by the 
large voids grow thicker as we increase the sampling density.
We see an extreme amount of fragmentation 
as the dark matter 
particles fill in structure marked by the large \emph{HOD} voids.
However, this fragmentation is limited to the edges of the voids, 
where the dark matter reveals additional substructure in the wall 
and filament networks surrounding each void; there is still a clear 
underdensity in the center. 
For medium-scale voids ($\sim 45$\hmpc), there is much 
less fragmentation in the dark matter.

\section{Relationships to matched voids}
\label{sec:matched}

We now directly compare voids in the two mock galaxy samples to their 
matched counterparts in the halo, low-resolution dark matter, and 
high-resolution dark matter populations. While this matching cannot 
be done in observations (where we do not have access to the dark 
matter distribution), it provides insight about the physical 
relationship of (simulated) galaxy and dark matter voids when reliable 
matches can be found.
We use the matching technique described in Section~\ref{sec:approach} 
for each pair of catalogs. Additionally, to assess the significance 
of our results we ran another simulation with identical cosmology 
and simulation parameters but with a different realization of the 
initial conditions. Thus for every matched void we also 
have a matched void from the alternate simulation.

We are unable to find matches for most of the smallest \emph{HOD Sparse} voids 
to the halo or low-resolution dark matter voids.
Conversely, we are able to find matches for nearly all \emph{HOD Dense} 
voids at all scales.
We are always able to find a match between galaxy voids 
at either density to the high-density dark matter voids, since 
the higher resolution uncovers significant substructure, as 
we have seen above.
Unsurprisingly, for both galaxy samples the smallest voids 
share little common volume 
with matches and match just as readily to voids in an alternate 
realization.

\begin{figure*} 
  \centering 
  {\includegraphics[type=png,ext=.png,read=.png,width=0.48\textwidth]{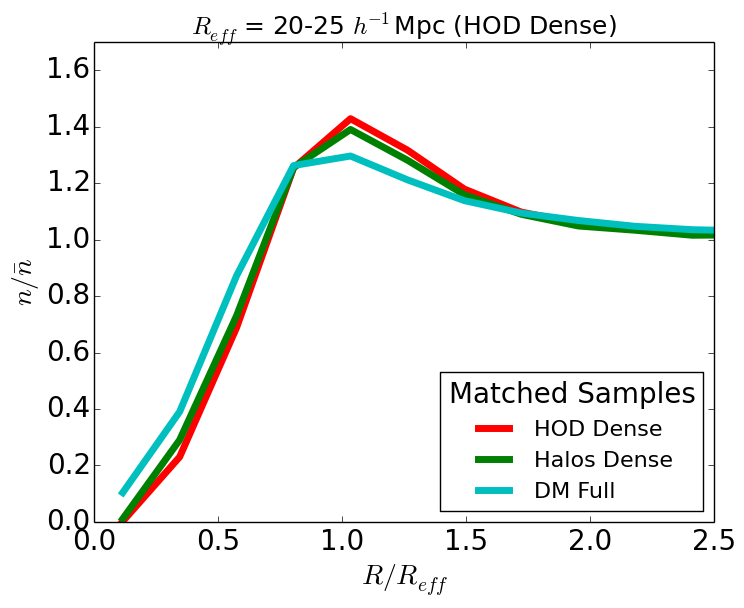}}
  {\includegraphics[type=png,ext=.png,read=.png,width=0.48\textwidth]{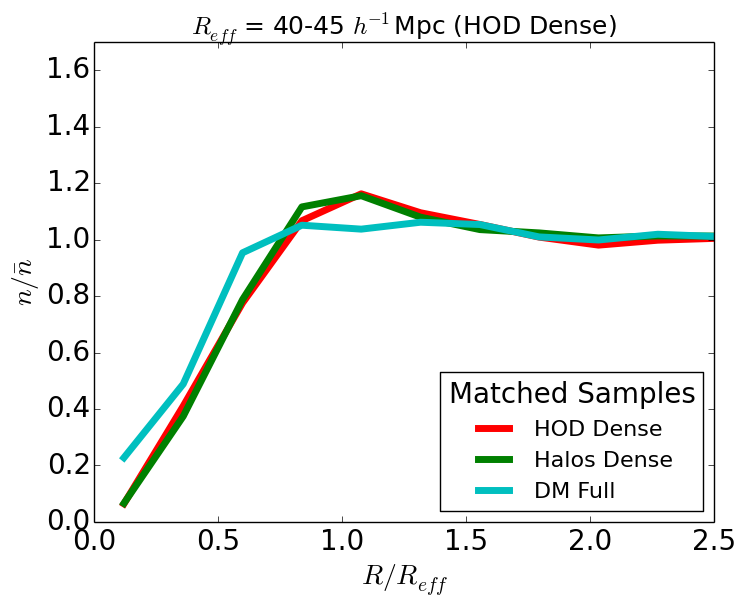}}
  {\includegraphics[type=png,ext=.png,read=.png,width=0.48\textwidth]{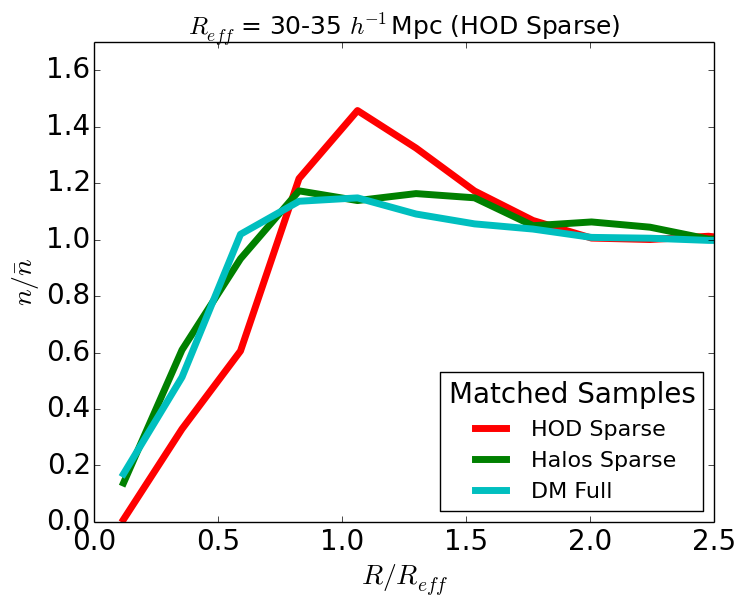}}
  {\includegraphics[type=png,ext=.png,read=.png,width=0.48\textwidth]{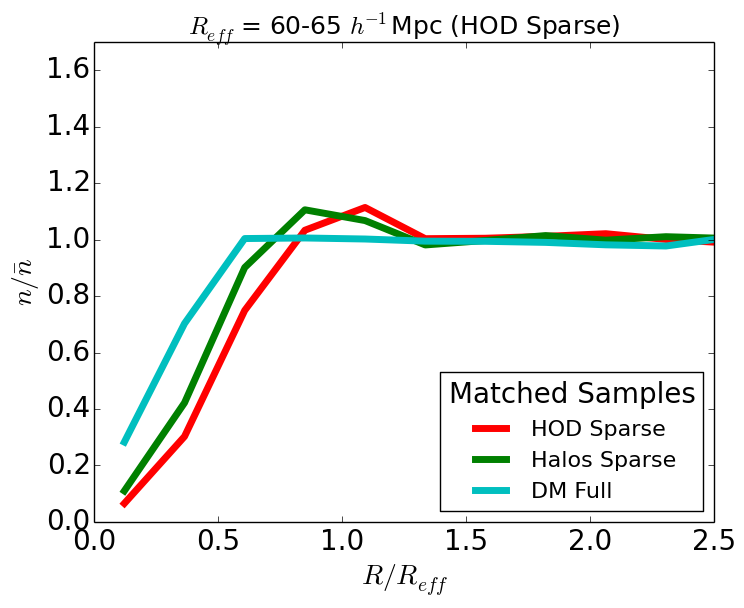}}
  \caption{
           Radial density profiles of stacked 
           voids in the \emph{HOD Dense} (top 
           row) and \emph{HOD Sparse} (bottom row) samples. For the 
           other samples listed, we build the profiles around the 
           centers of the best match voids to the galaxy voids
           in that stack, in contrast to Figure~\ref{fig:1dprofile_cocenter}
           where we keep the center fixed.
           We choose stacks to highlight overcompensated (left column) 
           and undercompensated (right column) void scales,
           and we indicate the void size ranges used in each stack at the 
           top of each plot. 
           We normalize each profile to the
           mean number density $\bar{n}$ of its corresponding sample.
         }
\label{fig:1dprofile_matched}
\end{figure*}

We construct the stacked radial profiles of 
Figure~\ref{fig:1dprofile_matched} in a similar fashion as in 
Figure~\ref{fig:1dprofile_cocenter}, with one major difference:
instead of keeping the centers of the stacks fixed on the galaxy 
voids, we shift the centers to the barycenter of each matched void 
in the stack. Thus, for each galaxy void in the given size range, 
we find the best match (based on volume overlap as described in 
Section~\ref{sec:approach}),
build the radial profile around that matched void, normalize 
the profile to the mean number density of that sample, and add that 
profile to the stack, regardless of its size. We do not rescale the 
voids as we add them to the stack.

In contrast with Figure~\ref{fig:1dprofile_cocenter}, these profiles 
lack any strong differences between the galaxy voids and the 
dark matter voids. While there is still some residual difference 
in the densities at the 
very central regions of each profile, and the compensation in the 
dark matter is not as high as in the galaxies, 
there are no significant changes to the slope. The profiles 
in the dark matter appear slightly broader and have lower 
compensations because we are including voids of a much 
wider radius range due to the matching (i.e., a 
single 25 \hmpc~galaxy void can 
match to voids of many sizes in the dark matter) than in the galaxy 
voids themselves. This broadening is not surprising: contrast, for example, the 
profiles from narrow stacks in~\citet{Sutter2012a} to the 
profile from all voids in~\citet{Pan2011}.
As with the profiles above, the \emph{HOD Dense} profiles 
are much more similar to the dark matter profiles than those from
the \emph{HOD Sparse} sample.

Since the matched voids only include voids inside the galaxy voids, 
these profiles show that each galaxy void corresponds to a deep 
underdensity in the dark matter, but that the centers are shifted.
Once this shift in position is taken into account, 
we can recover the expected universal 
density profile~\citep{LavauxGuilhem2011}.

The fraction of the volume of galaxy void shared between it and the matched 
voids tells a similar story.
Smaller voids in both galaxy samples 
share a significant fraction of their volume 
--- around 40\% --- with high-resolution dark matter voids, with 
larger voids sharing $\sim 20\%$ of their volume. 
For both galaxy mocks we see a clear distinction between matches 
to the same realization and matches to the alternate realization
for all halo and dark matter samples.
Indeed, we find that even though we are able to \emph{find} 
matches
in the alternate realization, and they may happen to be nearby 
the galaxy void, there is very little common volume,
even for the \emph{HOD Sparse} voids at all radii, meaning 
that there is still significant correspondence to the 
dark matter in the 
low-resolution galaxy voids.

\begin{figure*} 
  \centering 
  {\includegraphics[type=png,ext=.png,read=.png,width=0.48\textwidth]{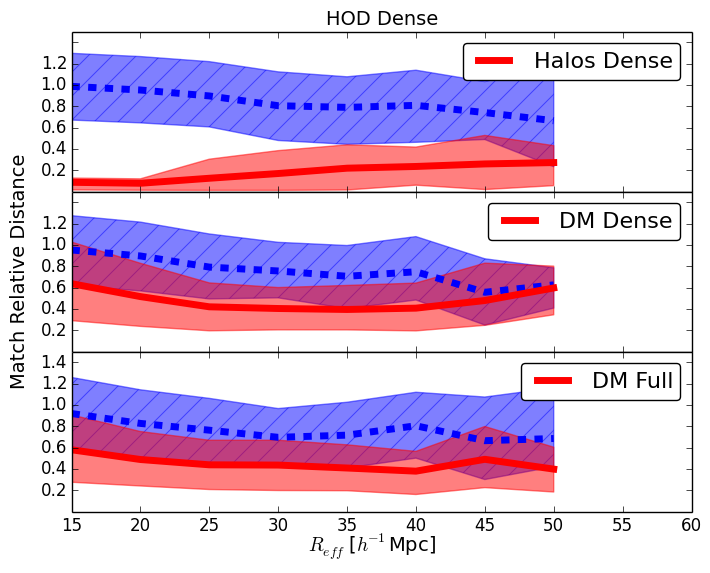}}
  {\includegraphics[type=png,ext=.png,read=.png,width=0.48\textwidth]{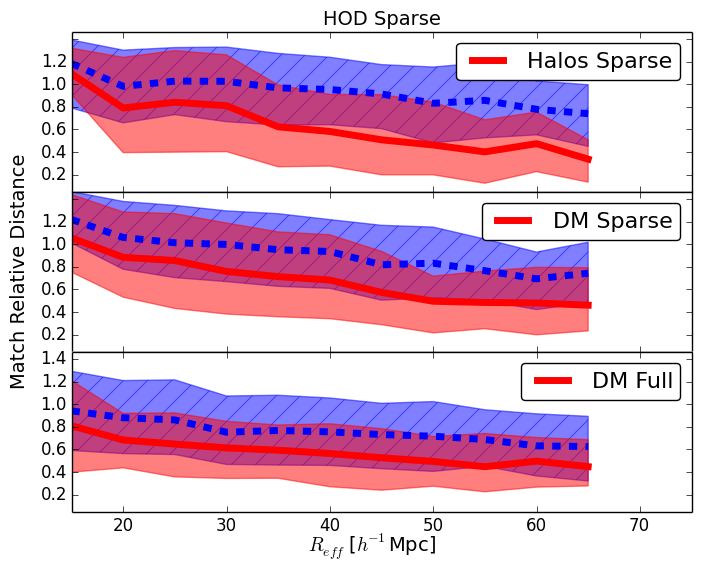}}
  \caption{
           Relative distance ($d/R_{\rm eff}$)
          between the centers of matched voids to the \emph{HOD Dense}
           (left) and \emph{HOD Sparse} (right) voids
           as a function of \emph{HOD} void effective radius.
           The solid lines indicate mean values in 5~\hmpc~bins 
           and shaded regions represent one standard deviation in 
           that bin. The red line and light red shaded area are 
           for matched to the same simulation, and the blue line and 
           light blue shaded area 
           are for matches to a simulation with a different realization 
           of an identical cosmology.
          }
\label{fig:matchvoldist}
\end{figure*}

Figure~\ref{fig:matchvoldist} shows the relative distance 
$d/R_{\rm eff}$ (where $R_{\rm eff}$ is the effective radius 
of the galaxy void) between the best-match void in each halo 
and dark matter sample and the \emph{HOD Dense} and \emph{HOD Sparse}
galaxy voids. We also plot the relative distance to the best match 
void in the alternate realization. 
Both the \emph{HOD Dense} and \emph{HOD Sparse} matches
follow similar trends: 
while some matched voids 
are no further than 0.3$R_{\rm eff}$, most have a relative distance 
of $\sim 0.6$, while a few of the smallest voids 
are further than the galaxy void effective radius (due to the 
ellipticity of the galaxy void). 
There are no significant differences between the distances to 
matches in the halo population and the dark matter population.
Finally, the mean distances to matches in the same realization are 
always shorter than matches to the alternate realization.
This suggests that there is a clear correspondence between 
galaxy and dark matter voids at almost all scales.

\begin{figure*} 
  \centering 
  {\includegraphics[type=png,ext=.png,read=.png,width=0.48\textwidth]{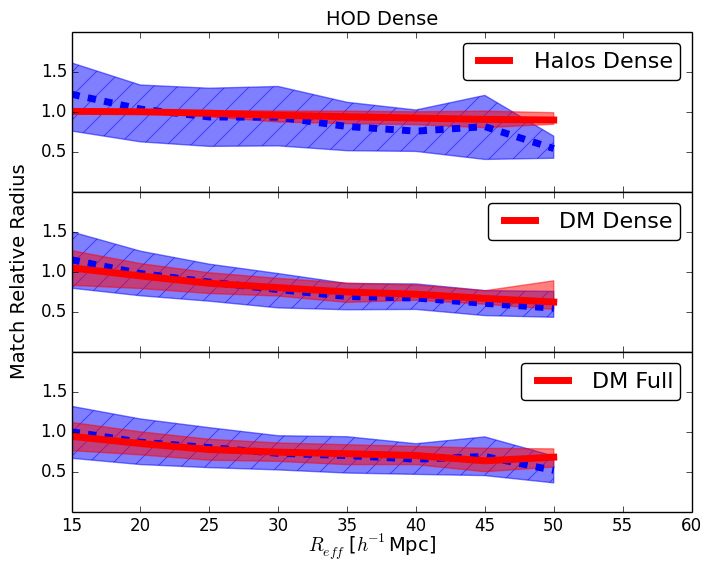}}
  {\includegraphics[type=png,ext=.png,read=.png,width=0.48\textwidth]{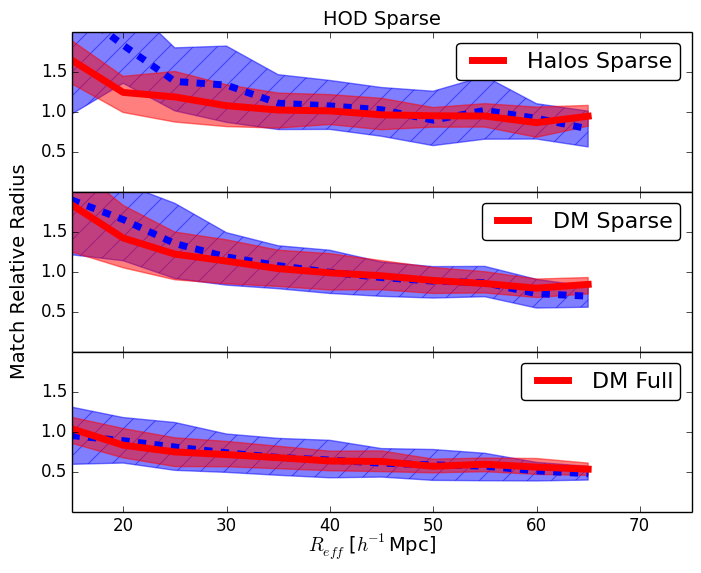}}
  \caption{
           Relative radius ($R_{\rm eff,match}/R_{\rm eff}$)
           between the centers of matched voids and the \emph{HOD Dense}
           (left) and \emph{HOD Sparse} (right) voids
           as a function of \emph{HOD} void effective radius.
           The solid lines indicate mean values in 5~\hmpc~bins 
           and shaded regions represent one standard deviation in 
           that bin. The red line and light red shaded area are 
           for matched to the same simulation, and the blue line and 
           light blue shaded area 
           are for matches to a simulation with a different realization 
           of an identical cosmology.
          }
\label{fig:matchvolrelradius}
\end{figure*}

In Figure~\ref{fig:matchvolrelradius} we show the relative size, 
which we define as $R_{\rm eff, match}/R_{\rm eff}$ of the 
void matched in the halo and dark matter populations to the 
galaxy voids. For both the \emph{HOD Sparse} and \emph{HOD Dense}
the match relative radius is roughly unity for the halos 
and low-resolution dark matter, and somewhat less than that 
(0.5 for \emph{HOD Sparse} and 0.75 for \emph{HOD Dense})
for the high-resolution dark matter. For both types of galaxy voids, 
the relative radius decreases for larger voids. 
This is because the larger voids in galaxies are more likely 
to contain substructure and more likely have wider walls in the dark matter, 
so the matched voids will necessarily be smaller.
There is only of order$\sim 10\%$ scatter in these relations 
for both galaxy samples.

\begin{figure} 
  {\includegraphics[type=png,ext=.png,read=.png,width=\columnwidth]{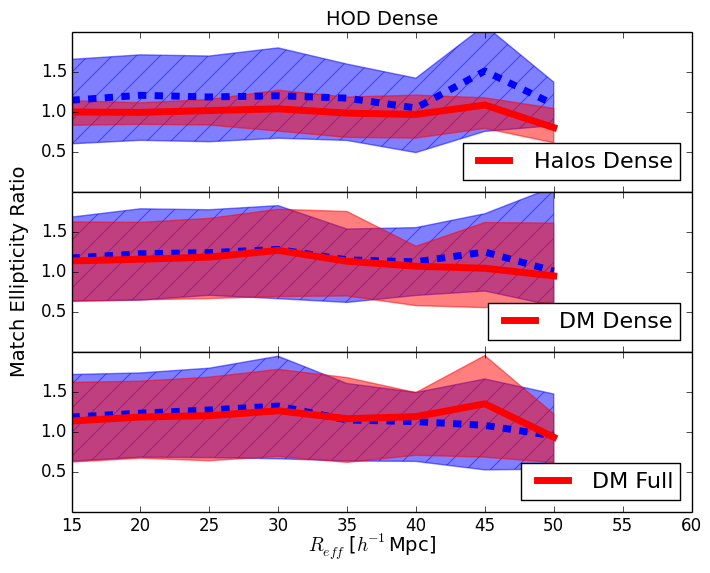}}
  \caption{
           Relative ellipticity ($\epsilon_{\rm match}/\epsilon$)
           between matched voids and the \emph{HOD Dense}
           voids
           as a function of \emph{HOD} void effective radius.
           The solid lines indicate mean values in 5~\hmpc~bins 
           and shaded regions represent one standard deviation in 
           that bin. The red line and light red shaded area are 
           for matched to the same simulation, and the blue line and 
           light blue shaded area 
           are for matches to a simulation with a different realization 
           of an identical cosmology.
           The \emph{HOD Sparse} voids show similar behavior and are 
           not shown.
          }
\label{fig:matchrelellip}
\end{figure}

However, the relationship between the shapes of the galaxy and matched 
voids is more complicated. 
To roughly measure the void shapes we compute the ellipticity.
For a given
set of tracers within a void we first construct the
inertia tensor:
\begin{eqnarray}
  M_{xx} & = &\sum_{i=1}^{N_p} (y_i^2 + z_i^2) \\ 
  M_{xy} & = & - \sum_{i=1}^{N_p} x_i y_i, \nonumber
\end{eqnarray}
where $N_p$ is the number of particles in the void, and
$x_i$, $y_i$, and $z_i$ are coordinates of the particle $i$
relative to the void barycenter.
The other components of the tensor are obtained by
cyclic permutations.
Given the inertia tensor, we compute the eigenvalues and form
the ellipticity:
\begin{equation}
  \epsilon = 1- \left( \frac{J_1}{J_3}\right)^{1/4},
\label{eq:ellip}
\end{equation}
where $J_1$ and $J_3$ are the smallest and largest eigenvalues,
respectively. 
Figure~\ref{fig:matchrelellip} shows the relative ellipticity 
between the galaxy voids and the halo and 
dark matter voids.

For both galaxy samples, the relative ellipticity is centered on 
unity for matches to voids in the halo populations,
suggesting that we are roughly 
capturing the void shape information with the galaxies. 
The mean relative ellipticity increases for matches to the dark matter 
voids, since generally voids are more elliptical in dark matter 
than galaxies.
However, there is also high variance in matches to the dark matter, and
there is no clear distinction between matches to 
the same realization and matches to the alternate realization.
This is not surprising, since ellipticity measurements 
of individual voids are very noisy due to the relatively few 
number of tracers within a void~\citep{Bos2012,Sutter2013a}, and 
small changes in the barycenter can greatly influence the 
resulting ellipticity measurement.
This emphasizes the importance of void \emph{stacking} 
to improve the overall signal-to-noise and obtain reliable 
shape information~\citep{LavauxGuilhem2011}. 

\section{Conclusions}
\label{sec:conclusions}

We have performed the most comprehensive analysis to date of the relationship
between voids in galaxy surveys and underdensities in dark matter. 
We have examined this relationship by building HOD mock galaxy
populations
within a high-resolution $N$-body simulation, finding voids 
in the galaxy distribution, and examining their radial density 
profiles of galaxies, halos, and dark matter.
We have further elucidated this relationship 
by matching each identified galaxy void to 
an individual dark matter void.
We have contrasted sparsely sampled 
and densely sampled galaxy surveys to asses the ability of current 
and future surveys to accurately identify and characterize voids 
in the cosmic web. We have also provided further 
examination of the common use of halo positions to represent a
galaxy survey, as previously studied in works 
such as~\citet{Padilla2005}.

Most importantly, we found that voids identified in both 
high- and low-resolution galaxy surveys
 correspond to physical underdensities in the dark 
matter. 
While the core densities in the corresponding dark matter tend to be 
higher than in the galaxy survey, at radii larger than roughly half 
the effective void radius the profiles are nearly identical.
Also, we have found that each galaxy void does contain a deep underdensity, 
but the location of that underdensity is usually 
offset from the galaxy void barycenter.
When this offset is taken 
into account, we recover a nearly universal density profile.
Measured at the void effective radius, mean densities in the 
galaxies are within 10\% of the mean densities in the dark matter. 

We have determined the capability of a particular survey to 
robustly identify given classes of voids. 
For high-resolution surveys, the smallest galaxy voids are most suspect, since 
they share little common volume with matches, tend to have highly 
displaced centers, and have large scatter in the relative radius 
to dark matter voids. They are also easily confused by matches 
to alternate realizations. 
However, above $\sim 25$-30 \hmpc, galaxy voids form a much 
tighter correspondence to dark matter voids. 
We also see ranges of void sizes in low-resolution surveys that 
appear reasonably well-matched to dark matter voids. 
While larger voids here tend to host significant dark matter 
substructure (i.e., fragmentation), this substructure is largely 
limited to the void edges; the interiors usually contain 
a single deep underdensity.
Medium-scale voids, in the range $\sim 25-55$ \hmpc, 
trace relatively the same structures as the dark matter.
These results confirm the statistical interpretation of~\citet{Hamaus2013}:
the cross-spectra of void and matter distributions indicate that 
void identified with the watershed approach do indeed inhabit 
underdense regions of the universe.

Voids at all scales in both high- and low-resolution surveys 
appear to map the same shapes as the dark matter voids, as 
we have seen in our ellipticity measurements. However, 
the noisiness of the ellipticity measurement leads to matches 
of the same significance to alternate realizations. Stacking procedures, 
which trade some loss of shape information for increased signal-to-noise 
and smoothening of the remaining shape, offer a promising 
alternative to this difficult situation~\citep{LavauxGuilhem2011, Sutter2012b}.

While we have utilized the matching procedure to gain insights into the 
relationship between galaxy voids and dark matter underdensities, 
we do not offer this as a prescription for inferring dark matter 
characteristics from an individual observed void. Instead, 
this analysis informs us on the relative uncertainty of 
void properties. For example, if we use the center of a galaxy 
void as a starting place for calculating the integrated 
Sachs-Wolfe effect~\citep{Planck2013b} or the lensing 
potential~\citep{Melchior2013}, 
and assume that this galaxy void corresponds to a single 
dark matter void, then this analysis
indicates that the
true center in the dark matter is likely within 
$\sim 50\%$ of the void radius.

We have targeted our mock galaxy populations to two specific surveys: 
the SDSS DR7 main sample ($M_r < -21$) 
and the SDSS DR9 CMASS sample, which we have 
taken to generally represent densely and sparsely sampled surveys, respectively.
Even though a full analysis should be undertaken to examine the quality of 
voids in future surveys such as BigBOSS~\citep{Schlegel2011} or 
Euclid~\citep{Laureijs2011}, 
this work provides some guidelines on the 
robustness of identified voids. 
While we have focused on the interior contents of voids in this paper, 
~\citet{Sutter2013a}
examines the statistical properties of galaxy void populations from 
realistic surveys.

We conclude that cosmological analyses that rest on the properties of voids
such as size distributions or radial profiles
are generally reliable when high-resolution galaxy surveys are used, 
as there is a clear correspondence between these voids and their 
dark matter counterparts. This correspondence is weakened for sparsely sampled surveys. However, it is not removed: there is a range of void scales 
that match reliably to dark matter.
Shape measurements with voids appear problematic in all surveys, 
but there are avenues available to address this such as statistical shape measurements in stacks of voids.
Most importantly, galaxy voids at all scales in all kinds of surveys 
still remain as underdensities 
in the dark matter and can still be considered as voids in the general 
sense. This is vitally important, since voids identified in galaxy surveys 
are a potentially rich source of astrophysical and cosmological information.

\section*{Acknowledgments}

The authors would like to thank Nico Hamaus for useful comments.
PMS and BDW acknowledge
support from NSF Grant AST-0908902. BDW
acknowledges funding from an ANR Chaire d’Excellence,
the UPMC Chaire Internationale in Theoretical Cosmology, and NSF grants AST-0908902 and AST-0708849. 
GL acknowledges support from CITA National Fellowship and financial
support from the Government of Canada Post-Doctoral Research Fellowship.
Research at Perimeter Institute is supported by the Government of Canada
through Industry Canada
 and by the Province of Ontario through the Ministry of Research and
Innovation.
DW acknowledges support from NSF Grant AST-1009505.

\footnotesize{
  \bibliographystyle{mn2e}
  \bibliography{voidinteriors}		

\begin{thebibliography}{}

\bibitem[\protect\citeauthoryear{{Ahn} et~al.,}{{Ahn}  et~al.}{2012}]{Ahn2012}
{Ahn} C.~P.,  et~al., 2012, \apjs, 203, 21

\bibitem[\protect\citeauthoryear{Alcock \& Paczynski}{Alcock \&
  Paczynski}{1979}]{Alcock1979}
Alcock C.,  Paczynski B.,  1979, Nature, 281, 358

\bibitem[\protect\citeauthoryear{{Aragon-Calvo} \& {Szalay}}{{Aragon-Calvo} \&
  {Szalay}}{2012}]{Aragon2012}
{Aragon-Calvo} M.~A.,  {Szalay} A.~S.,  2012, ArXiv e-prints: 1203.0248

\bibitem[\protect\citeauthoryear{{Bassett} \& {Hlozek}}{{Bassett} \&
  {Hlozek}}{2010}]{Bassett2010}
{Bassett} B.,  {Hlozek} R.,  2010, {Baryon acoustic oscillations}.
p.~246

\bibitem[\protect\citeauthoryear{{Baugh}}{{Baugh}}{2013}]{Baugh2013}
{Baugh} C.~M.,  2013, Publications of the Astronomical Society of Australia,
  30, 30

\bibitem[\protect\citeauthoryear{Behroozi, Wechsler \& Wu}{Behroozi
  et~al.}{2013}]{behroozi13}
Behroozi P.~S.,  Wechsler R.~H.,    Wu H.,  2013, The Astrophysical Journal,
  762, 109

\bibitem[\protect\citeauthoryear{Benson, Hoyle, Torres \& Vogeley}{Benson
  et~al.}{2003}]{Benson2003}
Benson A.~J.,  Hoyle F.,  Torres F.,    Vogeley M.~S.,  2003, \mnras, 340, 160

\bibitem[\protect\citeauthoryear{{Berlind} \& {Weinberg}}{{Berlind} \&
  {Weinberg}}{2002}]{Berlind2002}
{Berlind} A.~A.,  {Weinberg} D.~H.,  2002, \apj, 575, 587

\bibitem[\protect\citeauthoryear{Blas, Lesgourgues \& Tram}{Blas
  et~al.}{2011}]{blas11}
Blas D.,  Lesgourgues J.,    Tram T.,  2011, Journal of Cosmology and
  Astroparticle Physics, 2011, 034

\bibitem[\protect\citeauthoryear{{Bos}, {van de Weygaert}, {Dolag} \&
  {Pettorino}}{{Bos} et~al.}{2012}]{Bos2012}
{Bos} E.~G.~P.,  {van de Weygaert} R.,  {Dolag} K.,    {Pettorino} V.,  2012,
  ArXiv e-prints: 1205.4238

\bibitem[\protect\citeauthoryear{Ceccarelli, Padilla, Valotto \&
  Lambas}{Ceccarelli et~al.}{2006}]{Ceccarelli2006}
Ceccarelli L.,  Padilla N.~D.,  Valotto C.,    Lambas D.~G.,  2006, \mnras,
  373, 1440

\bibitem[\protect\citeauthoryear{{Clampitt}, {Cai} \& {Li}}{{Clampitt}
  et~al.}{2013}]{Clampitt2013}
{Clampitt} J.,  {Cai} Y.-C.,    {Li} B.,  2013, \mnras, 431, 749

\bibitem[\protect\citeauthoryear{Crocce, Pueblas \& Scoccimarro}{Crocce
  et~al.}{2006}]{crocce06}
Crocce M.,  Pueblas S.,    Scoccimarro R.,  2006, Monthly Notices of the Royal
  Astronomical Society, 373, 369{\textendash}381

\bibitem[\protect\citeauthoryear{Dehnen}{Dehnen}{2001}]{dehnen01}
Dehnen W.,  2001, Monthly Notices of the Royal Astronomical Society, 324,
  273{\textendash}291

\bibitem[\protect\citeauthoryear{Goldberg \& Vogeley}{Goldberg \&
  Vogeley}{2004}]{Goldberg2004}
Goldberg D.~M.,  Vogeley M.~S.,  2004, \apj, 605, 1

\bibitem[\protect\citeauthoryear{Gottlober, Lokas, Klypin \& Hoffman}{Gottlober
  et~al.}{2003}]{Gottlober2003}
Gottlober S.,  Lokas E.~L.,  Klypin A.,    Hoffman Y.,  2003, \mnras, 344, 715

\bibitem[\protect\citeauthoryear{{Gregory} \& {Thompson}}{{Gregory} \&
  {Thompson}}{1978}]{Gregory1978}
{Gregory} S.~A.,  {Thompson} L.~A.,  1978, \apj, 222, 784

\bibitem[\protect\citeauthoryear{{Hamaus}, Sutter, {Lavaux}, {Wandelt} \&
  Warren}{{Hamaus} et~al.}{2013}]{Hamaus2013}
{Hamaus} N.,  Sutter P.,  {Lavaux} G.,  {Wandelt} B.~D.,    Warren M.,  2013,
  \mnras (submitted)

\bibitem[\protect\citeauthoryear{Hoffman \& Shaham}{Hoffman \&
  Shaham}{1982}]{Hoffman1982}
Hoffman Y.,  Shaham J.,  1982, \apj, 262, L23

\bibitem[\protect\citeauthoryear{Hoyle \& Vogeley}{Hoyle \&
  Vogeley}{2004}]{Hoyle2004}
Hoyle F.,  Vogeley M.~S.,  2004, \apj, 607, 751

\bibitem[\protect\citeauthoryear{Komatsu et~al.,}{Komatsu
  et~al.}{2011}]{Komatsu2011}
Komatsu E.,  et~al., 2011, \apjs, 192, 18

\bibitem[\protect\citeauthoryear{{Laureijs} et~al.,}{{Laureijs}
  et~al.}{2011}]{Laureijs2011}
{Laureijs} R.,  et~al., 2011, {Euclid Definition Study Report}, arXiv:
  1110.3193

\bibitem[\protect\citeauthoryear{{Lavaux} \& {Wandelt}}{{Lavaux} \&
  {Wandelt}}{2011}]{LavauxGuilhem2011}
{Lavaux} G.,  {Wandelt} B.~D.,  2011, eprint arXiv:1110.0345

\bibitem[\protect\citeauthoryear{{Li}, {Zhao} \& {Koyama}}{{Li}
  et~al.}{2012}]{Li2012}
{Li} B.,  {Zhao} G.-B.,    {Koyama} K.,  2012, \mnras, 421, 3481

\bibitem[\protect\citeauthoryear{Li \& Zhao}{Li \& Zhao}{2009}]{Li2009}
Li B.,  Zhao H.,  2009, \prd, 80

\bibitem[\protect\citeauthoryear{{Little} \& {Weinberg}}{{Little} \&
  {Weinberg}}{1994}]{Little1994}
{Little} B.,  {Weinberg} D.~H.,  1994, \mnras, 267, 605

\bibitem[\protect\citeauthoryear{{Manera} et~al.,}{{Manera}
  et~al.}{2013}]{Manera2013}
{Manera} M.,  et~al., 2013, \mnras, 428, 1036

\bibitem[\protect\citeauthoryear{{Mar{\'{\i}}n} et~al.,}{{Mar{\'{\i}}n}
  et~al.}{2013}]{Marin2013}
{Mar{\'{\i}}n} F.~A.,  et~al., 2013, \mnras, 432, 2654

\bibitem[\protect\citeauthoryear{{Melchior}, {Sutter}, {Sheldon}, {Krause} \&
  {Wandelt}}{{Melchior} et~al.}{2013}]{Melchior2013}
{Melchior} P.,  {Sutter} P.~M.,  {Sheldon} E.~S.,  {Krause} E.,    {Wandelt}
  B.~D.,  2013, ArXiv e-prints: 1309.2045

\bibitem[\protect\citeauthoryear{Neyrinck}{Neyrinck}{2008}]{Neyrinck2008}
Neyrinck M.~C.,  2008, \mnras, 386, 2101

\bibitem[\protect\citeauthoryear{{Padilla}, {Ceccarelli} \& {Lambas}}{{Padilla}
  et~al.}{2005}]{Padilla2005}
{Padilla} N.~D.,  {Ceccarelli} L.,    {Lambas} D.~G.,  2005, \mnras, 363, 977

\bibitem[\protect\citeauthoryear{{Pan}, {Vogeley}, {Hoyle}, {Choi} \&
  {Park}}{{Pan} et~al.}{2012}]{Pan2011}
{Pan} D.~C.,  {Vogeley} M.~S.,  {Hoyle} F.,  {Choi} Y.-Y.,    {Park} C.,  2012,
  \mnras, 421, 926

\bibitem[\protect\citeauthoryear{{Parkinson} et~al.,}{{Parkinson}
  et~al.}{2012}]{Parkinson2012}
{Parkinson} D.,  et~al., 2012, \prd, 86, 103518

\bibitem[\protect\citeauthoryear{{Planck Collaboration}}{{Planck
  Collaboration}}{2013}]{Planck2013b}
{Planck Collaboration} 2013, ArXiv e-prints: 1303.5079

\bibitem[\protect\citeauthoryear{Platen, van~de Weygaert \& Jones}{Platen
  et~al.}{2007}]{Platen2007}
Platen E.,  van~de Weygaert R.,    Jones B. J.~T.,  2007, \mnras, 380, 551

\bibitem[\protect\citeauthoryear{Quinn, Katz, Stadel \& Lake}{Quinn
  et~al.}{1997}]{quinn97}
Quinn T.,  Katz N.,  Stadel J.,    Lake G.,  1997, {arXiv:astro-ph/9710043}

\bibitem[\protect\citeauthoryear{Ryden}{Ryden}{1995}]{Ryden1995}
Ryden B.~S.,  1995, \apj, 452, 25

\bibitem[\protect\citeauthoryear{{S{\'a}nchez} et~al.,}{{S{\'a}nchez}
  et~al.}{2012}]{Sanchez2012}
{S{\'a}nchez} A.~G.,  et~al., 2012, \mnras, 425, 415

\bibitem[\protect\citeauthoryear{{Schlegel} et~al.,}{{Schlegel}
  et~al.}{2011}]{Schlegel2011}
{Schlegel} D.,  et~al., 2011, {The BigBOSS Experiment}, arXiv:1106.1706

\bibitem[\protect\citeauthoryear{Schmidt, Ryden \& Melott}{Schmidt
  et~al.}{2001}]{Schmidt2001}
Schmidt J.~D.,  Ryden B.~S.,    Melott A.~L.,  2001, \apj, 546, 609

\bibitem[\protect\citeauthoryear{{Spolyar}, {Sahl{\'e}n} \& {Silk}}{{Spolyar}
  et~al.}{2013}]{Spolyar2013}
{Spolyar} D.,  {Sahl{\'e}n} M.,    {Silk} J.,  2013, ArXiv e-prints: 1304.5239

\bibitem[\protect\citeauthoryear{{Strauss} et~al.,}{{Strauss}
  et~al.}{2002}]{Strauss2002}
{Strauss} M.~A.,  et~al., 2002, \aj, 124, 1810

\bibitem[\protect\citeauthoryear{{Sutter}, {Lavaux}, {Wandelt}, {Hamaus},
  {Weinberg} \& {Warren}}{{Sutter} et~al.}{2013}]{Sutter2013a}
{Sutter} P.~M.,  {Lavaux} G.,  {Wandelt} B.~D.,  {Hamaus} N.,  {Weinberg}
  D.~H.,    {Warren} M.~S.,  2013, ArXiv e-prints: 1309.5087

\bibitem[\protect\citeauthoryear{{Sutter}, {Lavaux}, {Wandelt} \&
  {Weinberg}}{{Sutter} et~al.}{2012a}]{Sutter2012b}
{Sutter} P.~M.,  {Lavaux} G.,  {Wandelt} B.~D.,    {Weinberg} D.~H.,  2012a,
  \apj, 761, 187

\bibitem[\protect\citeauthoryear{{Sutter}, {Lavaux}, {Wandelt} \&
  {Weinberg}}{{Sutter} et~al.}{2012b}]{Sutter2012a}
{Sutter} P.~M.,  {Lavaux} G.,  {Wandelt} B.~D.,    {Weinberg} D.~H.,  2012b,
  \apj, 761, 44

\bibitem[\protect\citeauthoryear{{Sutter}, {Lavaux}, {Wandelt}, {Weinberg} \&
  {Warren}}{{Sutter} et~al.}{2013}]{Sutter2013c}
{Sutter} P.~M.,  {Lavaux} G.,  {Wandelt} B.~D.,  {Weinberg} D.~H.,    {Warren}
  M.~S.,  2013, ArXiv e-prints: 1310.7155

\bibitem[\protect\citeauthoryear{{Tavasoli}, {Vasei} \& {Mohayaee}}{{Tavasoli}
  et~al.}{2013}]{Tavasoli2013}
{Tavasoli} S.,  {Vasei} K.,    {Mohayaee} R.,  2013, \aap, 553, A15

\bibitem[\protect\citeauthoryear{{Tinker}, {Weinberg} \& {Zheng}}{{Tinker}
  et~al.}{2006}]{Tinker2006}
{Tinker} J.~L.,  {Weinberg} D.~H.,    {Zheng} Z.,  2006, \mnras, 368, 85

\bibitem[\protect\citeauthoryear{Warren}{Warren}{2013}]{warren13}
Warren M.~S.,  2013, in Proceedings of SC13: International Conference for High
  Performance Computing, Networking, Storage and Analysis SC '13, 2hot: An
  improved parallel hashed oct-tree n-body algorithm for cosmological
  simulation.
ACM, New York, NY, USA, pp 72:1--72:12

\bibitem[\protect\citeauthoryear{{Weinberg}, {Mortonson}, {Eisenstein},
  {Hirata}, {Riess} \& {Rozo}}{{Weinberg} et~al.}{2013}]{Weinberg2012}
{Weinberg} D.~H.,  {Mortonson} M.~J.,  {Eisenstein} D.~J.,  {Hirata} C.,
  {Riess} A.~G.,    {Rozo} E.,  2013, Phys. Rep., 530, 87

\bibitem[\protect\citeauthoryear{{Zehavi} et~al.,}{{Zehavi}
  et~al.}{2011}]{Zehavi2011}
{Zehavi} I.,  et~al., 2011, \apj, 736, 59

\bibitem[\protect\citeauthoryear{{Zheng}, {Coil} \& {Zehavi}}{{Zheng}
  et~al.}{2007}]{Zheng2007}
{Zheng} Z.,  {Coil} A.~L.,    {Zehavi} I.,  2007, \apj, 667, 760

\end{thebibliography}
}

\end{document}